\documentclass[journal]{IEEEtran}
\usepackage{tikz}
\usetikzlibrary{calc}

\ifCLASSINFOpdf
\else
\fi
\newcommand{\bvec}[1]{\ensuremath{{\boldsymbol{#1}}}}

\usepackage{mathtools}
\DeclarePairedDelimiterX{\infdivx}[2]{(}{)}{%
  #1\;\delimsize\|\;#2%
}

\newcommand{\Lshape}{\ensuremath{\mathsf{L}}-shape}
\usepackage{xcolor}

\newcommand{\Lhalf}{\ensuremath{L_{1/2}}}

\DeclareMathOperator*{\argmin}{\arg\!\min}

\def\rcurs{{\mbox{$\resizebox{.09in}{.08in}{\includegraphics[trim= 1em 0 14em 0,clip]{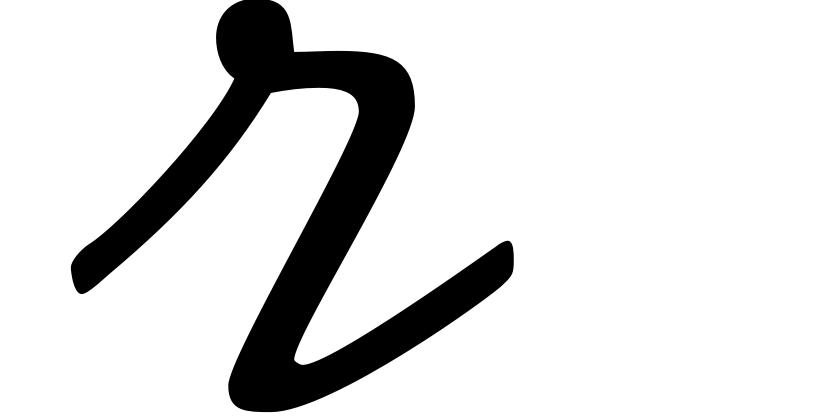}}$}}}

\usepackage{enumitem}

\usepackage{physics}
\usepackage{amsmath}
\usepackage{amssymb}
\usepackage{url}
\usepackage{hyperref}

\begin{document}

%
\title{Surrogate distributed radiological sources III: quantitative distributed source reconstructions}
%
%
%
\author{
    Jayson~R.~Vavrek,
    Jaewon~Lee,
    Marco~Salathe,
    Mark~S.~Bandstra,
    Daniel~Hellfeld,
    Brian~J.~Quiter,
    and Tenzing~H.Y.~Joshi
    \thanks{
        JR~Vavrek, M~Salathe, MS~Bandstra, D~Hellfeld, BJ~Quiter, and THY~Joshi are with the Applied Nuclear Physics program at Lawrence Berkeley National Laboratory.
        J~Lee is with the Department of Nuclear Engineering at UC Berkeley.
        This material is based upon work supported by the Defense Threat Reduction Agency under HDTRA 13081-36239.
        This support does not constitute an express or implied endorsement on the part of the United States Government.
        Distribution~A: approved for public release, distribution is unlimited.
    }%
}

\markboth{IEEE Transactions on Nuclear Science}%
{Vavrek \MakeLowercase{\textit{et al.}}}
%



\maketitle

\begin{abstract}
In this third part of a multi-paper series, we present quantitative image reconstruction results from aerial measurements of eight different surrogate distributed gamma-ray sources on flat terrain.
We show that our quantitative imaging methods can accurately reconstruct the expected shapes, and, after appropriate calibration, the absolute activity of the distributed sources.
We conduct several studies of imaging performance versus various measurement and reconstruction parameters, including detector altitude and raster pass spacing, data and modeling fidelity, and regularization type and strength.
The imaging quality performance is quantified using various quantitative image quality metrics.
Our results confirm the utility of point source arrays as surrogates for truly distributed radiological sources, and advance the quantitative capabilities of Scene Data Fusion gamma-ray imaging methods.

\end{abstract}


%
\IEEEpeerreviewmaketitle

\section{Introduction}
\IEEEPARstart{A}{s} described in Parts~I~\cite{vavrek2024surrogateI} and II~\cite{vavrek2024surrogateII} of this multi-paper series, we designed, fielded, and measured surrogate distributed radiological sources constructed from arrays of up to~$100$ Cu-64 point sources.
In this Part~III, we finally turn to the task of quantitatively imaging these distributed sources, i.e., determining both the shape and magnitude of radiation intensities in the mapped area.

Quantitative distributed source image reconstruction represents an advance in complexity compared to the two more common distributed source mapping modalities, namely either simply measuring dose rates at multiple points and interpolating between them (``breadcrumbing''), or performing only qualitative, relative ``hot'' vs.~``cold'' reconstructions with no absolute scale.
This increase in complexity, however, provides an additional wealth of information that may be used to drive decision-making, such as whether a radionuclide concentration in the environment is above regulatory limits, or how to minimize the dose to personnel at ground level based on aerial measurements.

In addition to Parts~I and II, this paper builds upon much of our previous work demonstrating free-moving quantitative gamma-ray imaging of small (${\sim}1$~m$^2$) distributed sources~\cite{hellfeld2021free}, and is part of a larger effort in advancing the quantitative capabilities of a radiation mapping technique we have termed ``Scene Data Fusion'' (SDF)~\cite{vetter2019advances}, which leverages contextual sensors to describe the image space to which measured radiation is attributed.
Distributed source mapping has also been of interest to several other groups concerned with radiological emergency response---see for instance Refs.~\cite{murtha2021tomographic, macleod20233, wahl20233d, daniel2021extended}.
This paper in particular offers a number of contributions over existing work.
First, while the state of the art often uses traditional Compton imagers or passive coded mask imagers, both of which typically have a limited field of view, this work focuses on singles (i.e., non-Compton) imaging with omnidirectional gamma-ray imagers.
Second, compared to the existing literature (which, again, often produces only qualitative, relative-scale images), the well-known configuration of the point source arrays allow us to make more thorough quantitative comparisons between the reconstructed images and the ground truth configurations.
Third, due to the reconfigurability of the array sources, we were able to measure a number of distributed source patterns of varying complexity, offering a more comprehensive study of various intensity and spatial features.
Finally, we also investigate quantitative reconstruction performance for a wide variety of experiment and reconstruction parameters.

The paper structure is as follows: Section~\ref{sec:methods} provides an overview of the measurement campaign, quantitative reconstruction methods used, and image quality metrics chosen.
Section~\ref{sec:results} then provides quantitative reconstruction results in a number of studies.
In particular, we show sample imaging results for the eight distributed source types using representative flight parameters and reconstruction hyper-parameters, and then perform sweeps over various (hyper-)parameters such as detector altitude and regularization strength.
Sections~\ref{sec:discussion} and~\ref{sec:conclusion} then conclude with a further discussion of results and a general summary of findings.

\section{Methods}\label{sec:methods}

\subsection{Measurement overview}
As described in Part~II~\cite{vavrek2024surrogateII}, in August 2021 we conducted a week-long experimental campaign at Washington State University (WSU) using radiation detectors borne on unmanned aerial systems (UASs) to measure the gamma-ray signature of various surrogate distributed Cu-64 source patterns.
The source patterns were constructed out of arrays of up to~$100$ point sources spaced densely enough to ``look like'' a truly continuous distributed source of Cu-64 for a given detector and trajectory (according to the method developed in Part~I, Section~II), and range in complexity from a uniform square source to regions of higher and zero activity superimposed on a uniform baseline (see Part~I, Fig.~3).
The nominal point source activities were ${\sim}7$~mCi at 0800~PDT of each measurement day, though the sources decayed throughout the day with a $12.7$-hour half-life, and the count rate comparisons made in Part~II Section~V-A suggest that the true source activities were ${\sim}1.4\times$ higher than the planned nominal activities, which were subject to substantial calibration uncertainties.
In this work, we consider this ${\sim}1.4\times$ scale factor to be part of the overall system calibration.
The predominant decay signature of Cu-64 is the $511$~keV annihilation photon line, which can be detected either in `singles' mode (i.e., through its photopeak) or in `doubles' mode (i.e., through its Compton scattering between two detector segments).
In this work, we restrict ourselves to singles imaging, but note that Compton imaging analyses remains of interest.
The detector systems used are NG-LAMP ($4 \times 1'' \times 1'' \times 2''$ CLLBC modules)~\cite{Pavlovsky2019} and MiniPRISM ($58 \times 1\,\text{cm} \times 1\,\text{cm} \times 1\,\text{cm}$ CZT modules)~\cite{pavlovsky2019miniprism}, both of which are capable of omnidirectional gamma-ray imaging.
Due to the imperfect energy resolution of real detectors, we define the energy region of interest (ROI) corresponding to the $511$~keV photopeak to be $[ 481, 581 ]$~keV for NG-LAMP and $[ 481, 541 ]$~keV for MiniPRISM---see Part~II\@.
Also as described in Part~II, when necessary we exclude data from detector crystals that regularly exhibited poor spectral and/or timing performance throughout the campaign.
As such, $2/4$ NG-LAMP crystals and $53/60$ MiniPRISM crystals were used for these analyses.
Finally, we restrict our analysis to data collected during the constant-altitude UAS raster patterns, i.e., we exclude takeoff/landing and any other free-form trajectories performed.

\subsection{Scene generation and co-registration}

To make detailed quantitative comparisons between the distributed sources designed in Part~I and their SDF reconstructions, accurate models of the measured scene must be generated and then aligned with the designed coordinate frame to as good a precision as possible.

\begin{figure*}[!htbp]
    \centering
    \includegraphics[width=0.51\linewidth]{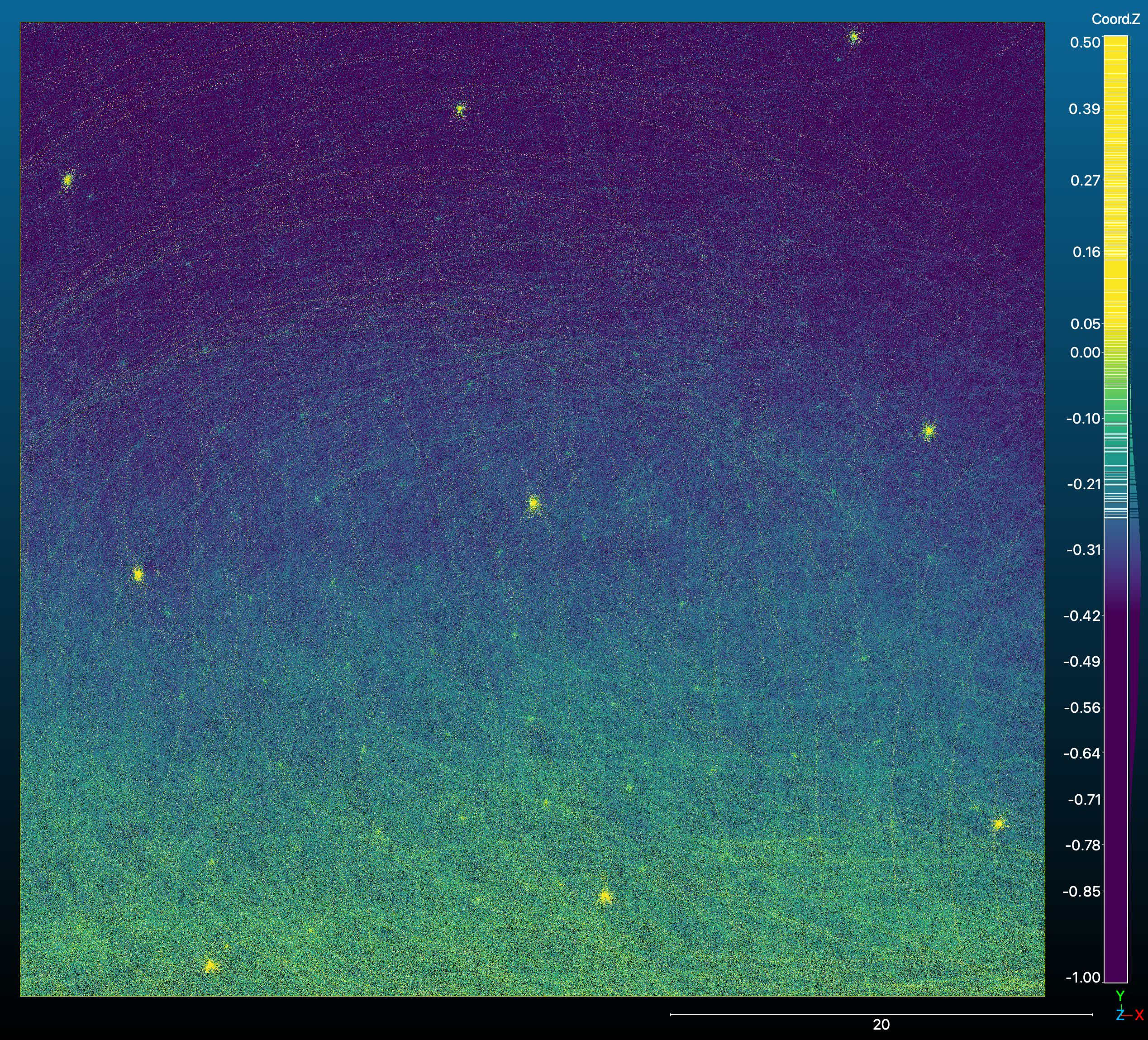}
    \includegraphics[width=0.48\linewidth]{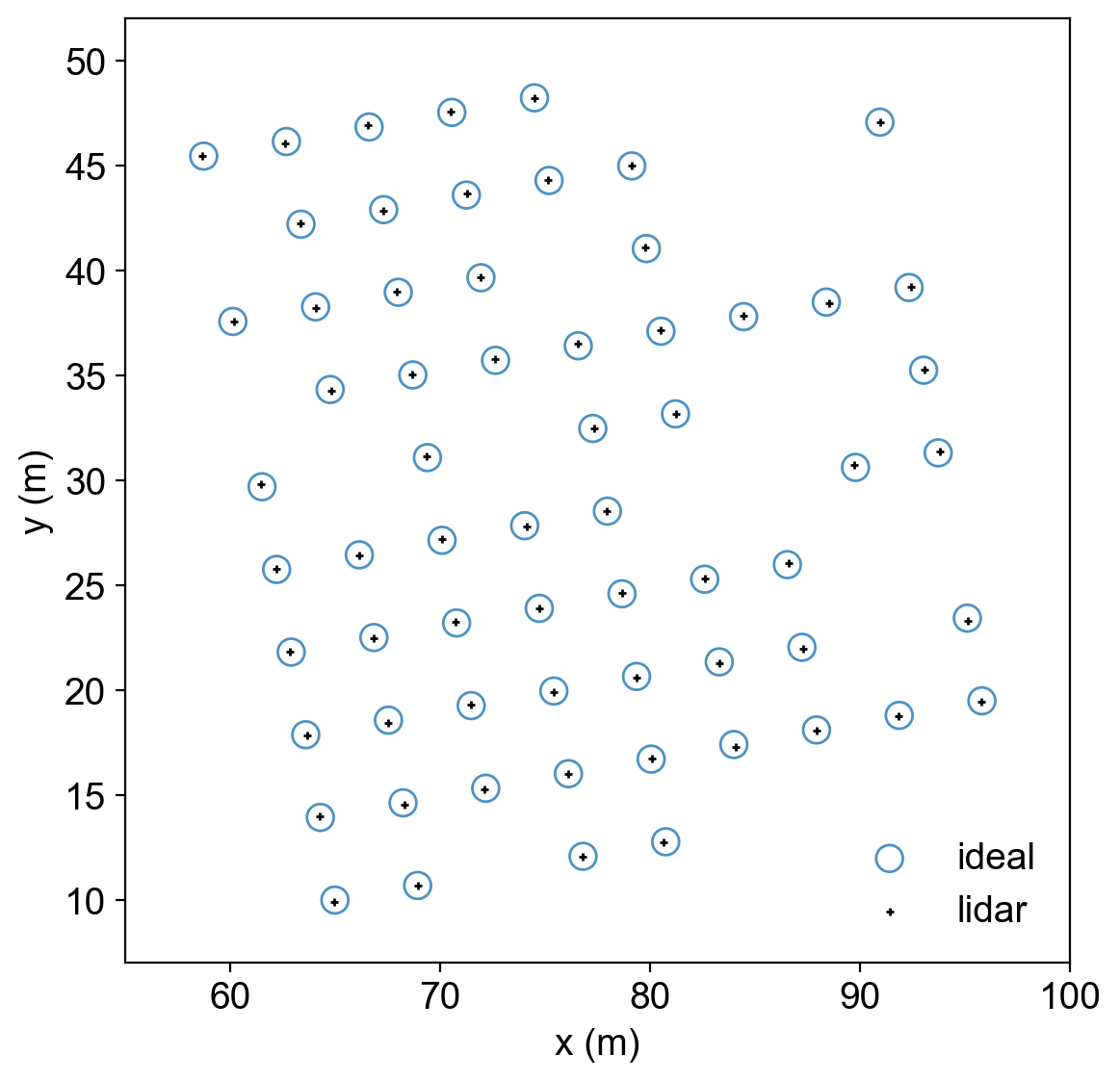}
    \caption{
        Left: top-down view of the (un-leveled) square source reference point cloud in CloudCompare, with points colorized by height.
        The larger yellow point clusters are traffic cones around the source boundary and at the $10 \times 10$ grid center, while the grid of smaller yellow point clusters corresponds to the individual source locations.
        Right: $xy$ positions of the square source locations discernible in the (leveled) reference point cloud (black~$+$) after least-squares alignment to their corresponding points in the designed square source (blue~$\circ$).
        The sizes of the markers are intended only to best show the point-to-point alignment rather than represent any uncertainties.
    }
    \label{fig:srcs_aligned_xy}
\end{figure*}

The NG-LAMP and MiniPRISM detector systems are each equipped with a lidar and inertial measurement unit (IMU) to enable lidar-based simultaneous localization and mapping (SLAM)~\cite{durrant2006simultaneous, bailey2006simultaneous} in order to reconstruct the detector trajectory and a 3D digital point cloud model of the mapped area.
To co-register all SLAM results to the idealized field coordinate frame used in Part~I, we first identified a high-quality SLAM map of a square source run in which most of the individual source locations were resolvable.
The field had a slope of ${\sim}0.5^\circ$ for drainage, so we then aligned the field normal to the $+z$ axis by fitting a small plane near the center of the square source.
After the point cloud was leveled, we were able to identify $64$ of the $100$ source locations through manual inspection in the CloudCompare software~\cite{girardeau2016cloudcompare}, and imputed a constant $z$ source position based on the apparent ground surface within the point cloud.
These lidar source position estimates were then aligned with their corresponding points in the ideal square source (of exactly $4$~m grid spacing) using a least-squares distance minimization between all $64$ corresponding point sets, and the resulting transformation was applied to the leveled point cloud.
The mean (absolute) distance in the $xy$ plane between the designed and lidar source points after co-registration was $6.2$~cm (with the mean signed $x$- and $y$-component differences of $\ll$~1~mm indicating no directional bias), and the mean grid spacing between the lidar source points was $4.0047$~m, only $0.12\%$ different from the design value of $4$~m.
Fig.~\ref{fig:srcs_aligned_xy} shows the lidar point cloud colorized by height and the $xy$ positions of the $64$ discernible square source locations after alignment.

\begin{figure*}[!htbp]
    \centering
    \includegraphics[width=0.9\linewidth]{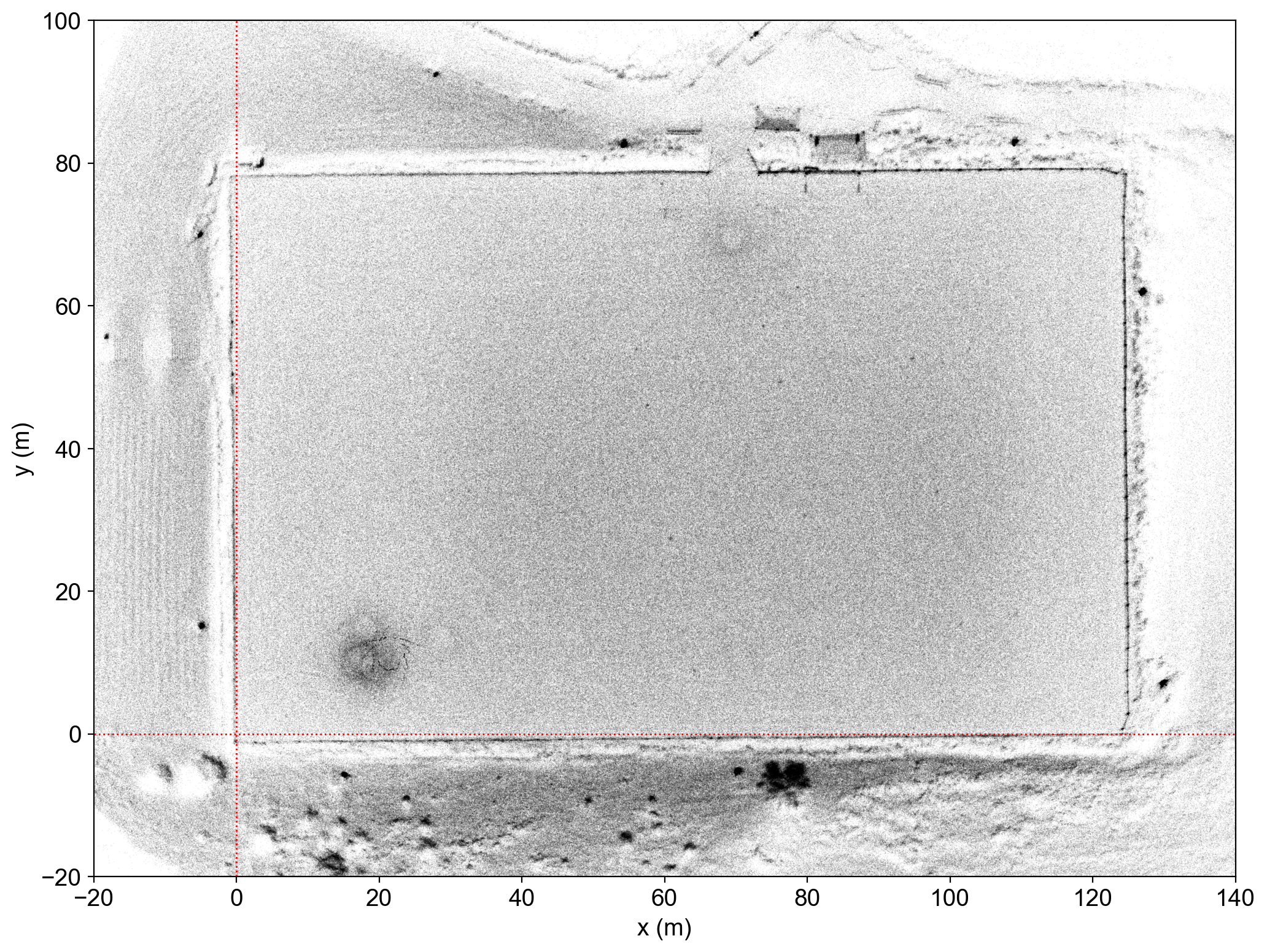}\\
    \includegraphics[width=0.9\linewidth]{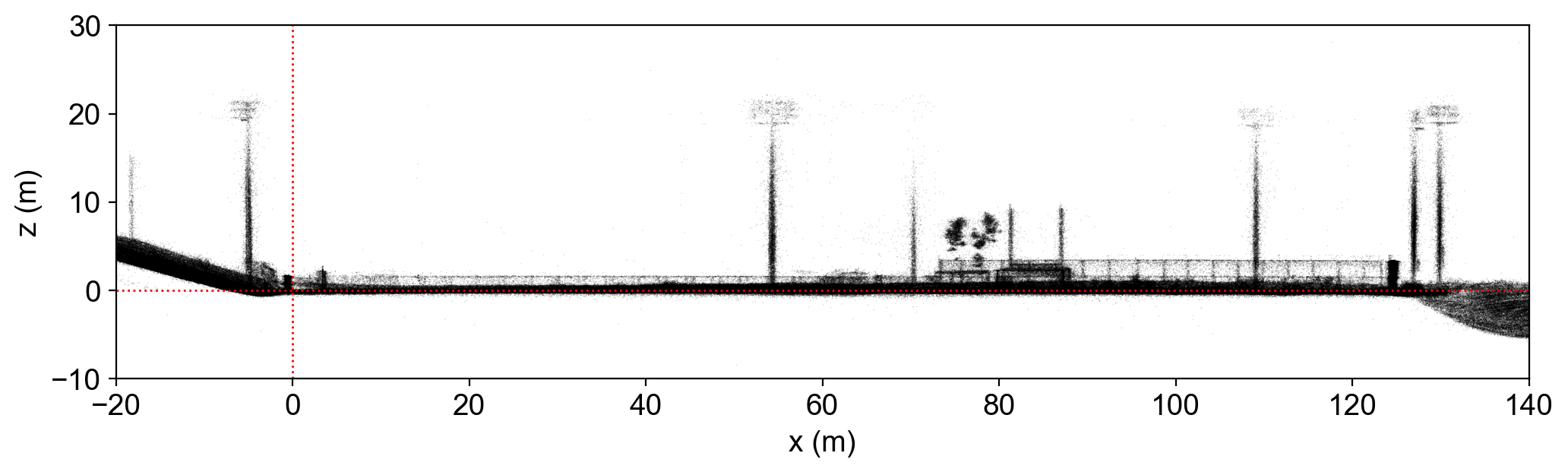}\\
    \includegraphics[width=0.9\linewidth]{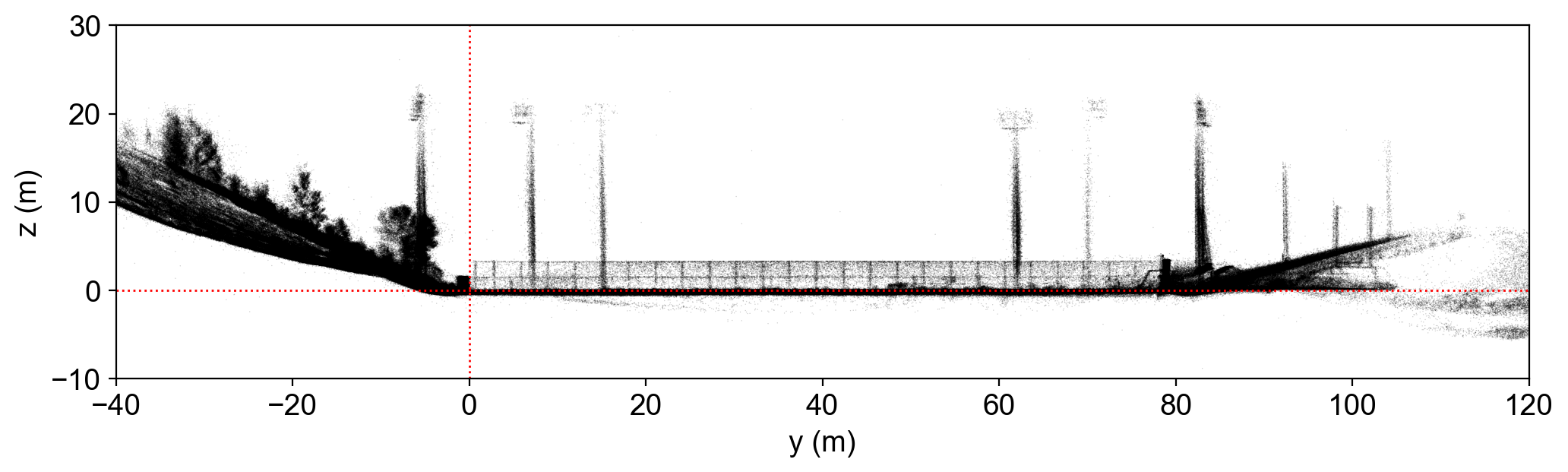}
    \caption{
        $35$ lidar SLAM point clouds from the WSU aerial measurement campaign, all co-registered and aligned to the idealized field coordinate system of Part~I\@.
        Top: $xy$~projection.
        The darker spots in the center-right of the field are traffic cones placed around the boundary of the $10\times 10$~square and \Lshape{} sources.
        The circular patterns near $(x,y)=(20,10)$ and $(70,75)$~m are lidar artifacts from takeoff and landing.
        Middle: $xz$~projection within the $y$ boundaries of the field.
        Bottom: $yz$~projection within the $x$ boundaries of the field.
    }
    \label{fig:clouds_aligned}
\end{figure*}

The remaining point clouds were then co-registered to the now-aligned reference cloud using the Iterative Closest Point (ICP) algorithm in Open3D~\cite{zhou2018open3d}.
We enforced a minimum ICP fitness score of $0.98$ (chosen empirically), and manually provided improved initial guesses for the transform if the initial ICP result did not meet this threshold on the first pass.
Fig.~\ref{fig:clouds_aligned} shows the results of this co-registration procedure on $35$ point clouds from the campaign.
Finally, after all measurements were co-registered, the volume for radiological source reconstruction was defined as the plane $x \in [0, 125] \cap y \in [0, 80]$~m at $z=0$, with a $1 \times 1$~m pixel size.

\subsection{Image reconstruction}
As described in Part~I Section~II-A, we perform single-energy gamma-ray imaging by assuming a linear model
\begin{align}\label{eq:imaging_model}
    \bvec{\lambda} = \boldsymbol{V} \bvec{w},
\end{align}
where $\bvec{w} \in \mathbb{R}^K_{+}$ is the vector of gamma-ray intensities or ``weights'' at each point or voxel center indexed $k=1, \ldots , K$, $\bvec{\lambda} \in \mathbb{R}^I_{+}$ is the vector of expected counts observed by a detector over a series of measurements indexed $i = 1, \ldots , I$, $\boldsymbol{V} \in \mathbb{R}^{I \times K}_{+}$ is the system matrix that relates the weights to the expected counts, and the $+$ subscript denotes non-negativity.
An unknown but constant-in-time background term, $b$, can also be included and solved for (see Ref.~\cite{hellfeld2021free}), but we exclude it in this discussion for clarity and because our measurements are heavily source-dominated.
The $(i, k)$ element of the system matrix is
\begin{align}\label{eq:sys_mat}
    V_{ik} = \frac{\eta_{ik} t_i}{4\pi | \vec{\rcurs}_{ik}|^2} \exp \left( -\mu_\text{air} |\vec{\rcurs}_{ik}| \right),
\end{align}
where $\eta_{ik}$ is the effective area to source point~$k$ during measurement~$i$, $\vec{\rcurs}_{ik}$ is the corresponding distance vector, and~$t_i$ is the $i\textsuperscript{th}$ dwell time.
The effective area incorporates both the geometric and intrinsic efficiency of the detector, including attenuation from other detector modules and components in the system---for example, see Fig.~5(a) of Ref.~\cite{hellfeld2021free}, which shows the MiniPRISM effective area as a function of angle to the source.
A measurement (or series thereof) then consists of a Poisson sample
\begin{align}
    \bvec{n} \sim \text{Poisson}(\bvec{\lambda}),
\end{align}
which has a corresponding Poisson negative log-likelihood of
\begin{align}\label{eq:pnll}
    \ell(\bvec{n} | \bvec{\lambda}) = \left[ \bvec{\lambda} - \bvec{n} \odot \log (\bvec{\lambda}) + \log(\bvec{n}!) \right] \cdot \bvec{1},
\end{align}
where $\odot$ denotes element-wise multiplication.
Eq.~\ref{eq:pnll} can also be used in defining the ``unit deviance'' or ``deviance residual'' between the data and the model, which reduces to a $\chi^2$~metric at high counts---see Part I, Section~II-D\@.

A maximum likelihood radiation image reconstruction algorithm then seeks the optimization
\begin{align}\label{eq:ml_def}
    \hat{\boldsymbol{w}} = \argmin_{\bvec{w} \ge 0} \, \ell(\bvec{n} | \bvec{\lambda}),
\end{align}
or, by incorporating prior information, a maximum \textit{a posteriori} algorithm seeks
\begin{align}\label{eq:map_def}
    \hat{\boldsymbol{w}} = \argmin_{\bvec{w} \ge 0} \, \left[ \ell(\bvec{n} | \bvec{\lambda}) + \beta f(\bvec{w}) \right]
\end{align}
for some hyperparameter $\beta > 0$ and regularization function $f(\bvec{w})$, which are often chosen to promote sparseness and/or smoothness in the reconstructed image $\hat{\boldsymbol{w}}$.
Eq.~\ref{eq:ml_def} can be solved iteratively through the maximum likelihood expectation maximization (ML-EM) algorithm~\cite{shepp1982maximum, lange1984reconstruction}
\begin{align}\label{eq:mlem}
    \hat{\bvec{w}}^{(m+1)} = \frac{\hat{\bvec{w}}^{(m)}}{\bvec{\varsigma}} \odot \boldsymbol{V}^\top \frac{\bvec{n}}{\boldsymbol{V} \hat{\bvec{w}}^{(m)}},
\end{align}
where $\hat{\bvec{w}}^{(0)}$ is typically initialized to a flat image and the sensitivity map
\begin{align}
    \boldsymbol{\varsigma} \equiv \boldsymbol{V}^\top \boldsymbol{1} \in \mathbb{R}^K_{+}
\end{align}
describes the expected number of counts per unit weight for each source element $k$ summed over the $I$ measurements.
Moreover Eq.~\ref{eq:mlem} can be extended to solve the regularized minimization in Eq.~\ref{eq:map_def} (``maximum \textit{a posteriori} expectation maximization'' or MAP-EM)---see Ref.~\cite{hellfeld2021free}, Eqs.~7--8 for further details.
In particular we consider the sparsity-promoting, non-convex \Lhalf{} prior~\cite{xu2010lhalf},
\begin{align}
    f(\boldsymbol{w}) \equiv \sqrt{\boldsymbol{w}} \cdot \boldsymbol{1},
\end{align}
and the smoothing and edge-preserving total variation (TV) prior~\cite{rudin1992nonlinear, panin1998total}, which can be written in 2D as
\begin{align}\label{eq:TV}
    f(\boldsymbol{w}) \equiv \iint \sqrt{ \left(\frac{\dd  \boldsymbol{w}}{\dd x}\right)^2 + \left(\frac{\dd  \boldsymbol{w}}{\dd y}\right)^2} \dd x\, \dd y.
\end{align}
Computationally, the integrals in Eq.~\ref{eq:TV} are replaced with discrete sums and the derivatives with finite differences.
An additional parameter $\varepsilon$ is often introduced to stabilize calculations when taking the gradient of Eq.~\ref{eq:TV}~\cite[\S II-C]{panin1998total}.
Reconstruction algorithms are implemented in the {\tt mfdf}~\cite{joshi2020mfdf} package using {\tt radkit} libraries~\cite{joshi2021radkit}, and run on a graphics processing unit (GPU) via PyOpenCL\@.

\subsection{Image quality metrics}
To evaluate the quality of a reconstructed radiation image~$\bvec{\hat{w}}$ compared to its corresponding ground truth image $\bvec{w}$, we will typically consider three metrics:
(1) the ratio of total reconstructed and true activities,
\begin{align}
    R_\text{tot}(\bvec{\hat{w}}, \bvec{w}) \equiv \frac{\bvec{\hat{w}} \cdot \boldsymbol{1}}{\bvec{w} \cdot \boldsymbol{1}} \equiv \frac{\hat{w}_\text{tot}}{w_\text{tot}},
\end{align}
(2) the normalized root-mean-square error (NRMSE),
\begin{align}
    \text{NRMSE}(\bvec{\hat{w}}, \bvec{w}) \equiv \frac{1}{w_\text{max} - w_\text{min}} \sqrt{\frac{(\bvec{\hat{w}} - \bvec{w})^2 \cdot \boldsymbol{1}}{K}},
\end{align}
and (3) the structure coefficient from Eq.~10 of Ref.~\cite{wang2004image},
\begin{align}
    s(\bvec{\hat{w}}, \bvec{w}) \equiv \frac{\text{cov}(\bvec{\hat{w}}, \bvec{w}) + \epsilon}{\sqrt{\text{var}(\bvec{\hat{w}}) \text{var}(\bvec{w})} + \epsilon},
\end{align}
for some small stabilizing constant $\epsilon > 0$.
We note that the structure coefficient,~$s$, is essentially the Pearson correlation coefficient between the image pixel intensities.
The activity ratio, $R_\text{tot}$, is chosen to give an overall intensity accuracy (closer to $1$ is better), the NRMSE is chosen to measure the average intensity similarity over the image independent of pixel position and normalized by the true range (closer to $0$ is better), and the structure metric,~$s$, is chosen to measure \textit{perceived} image similarity (closer to $1$ is better).
In Figs.~\ref{fig:imaging_study_low}, \ref{fig:imaging_study_high}, \ref{fig:altitude_study}--\ref{fig:speed_study}, and~\ref{fig:regularizer0_study}--\ref{fig:raster_spacing_study}, these metrics are displayed on each reconstructed activity image.

Because the on-field source distributions consisted of ${\leq}100$ individual point sources specifically configured to emulate continuous distributed sources, for the purposes of quantitative image comparison the ``true'' images $\bvec{w}$ are generated by interpolating the designed point source arrays to $1$-m-pixel continuous ground truth images with the same average activity concentrations.

While reconstructed images are computed over the entire $125\, \text{m} \times 80\, \text{m}$ field, much of that area has zero source present, and, as shown in Section~\ref{sec:results}, correspondingly quite low reconstructed activity compared to the areas with source present.
Thus, to reduce the extent to which image quality metrics are dominated by near-zero comparisons, we compute the structure and NRMSE metrics over a smaller subset of the image space, namely $x \in [40, 120] \cap y \in [0, 80]$~m.

\section{Results}\label{sec:results}

\subsection{Image study}
\begin{figure*}[!htbp]
    \centering
    \includegraphics[width=2.0\columnwidth]{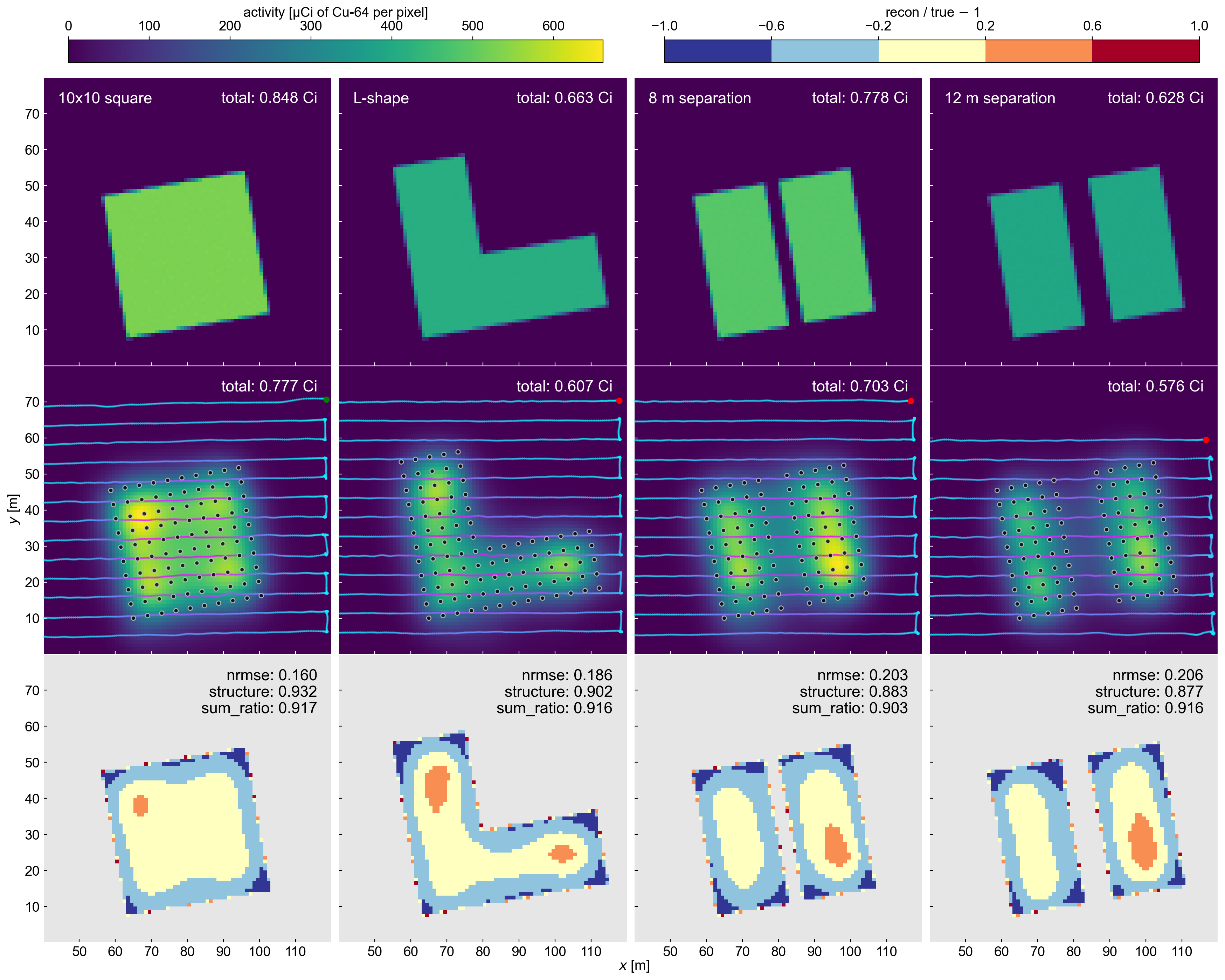}
    \caption{
        Imaging study for four surrogate distributed sources (left to right: the $10 \times 10$ square, the \Lshape{}, the $8$~m separation, and the $12$~m separation sources) using the NG-LAMP system.
        Top row: ground truth source distributions interpolated to the $1 \times 1$~m-pitch reconstruction grid.
        Middle row: MAP-EM reconstructed source distributions, array source distributions, and detector trajectories derived from SLAM\@.
        Bottom row: differences between the reconstructed and interpolated true distributions, normalized by the latter.
        Gray pixels correspond to true source activities of~$0$.
        The colorbar on the left corresponds to the source distributions (top row) and the reconstructed activities (middle row) while the right-hand colorbar corresponds to the bottom row relative difference maps.
    }
    \label{fig:imaging_study_low}
\end{figure*}

\begin{figure*}[!htbp]
    \centering
    \includegraphics[width=2.0\columnwidth]{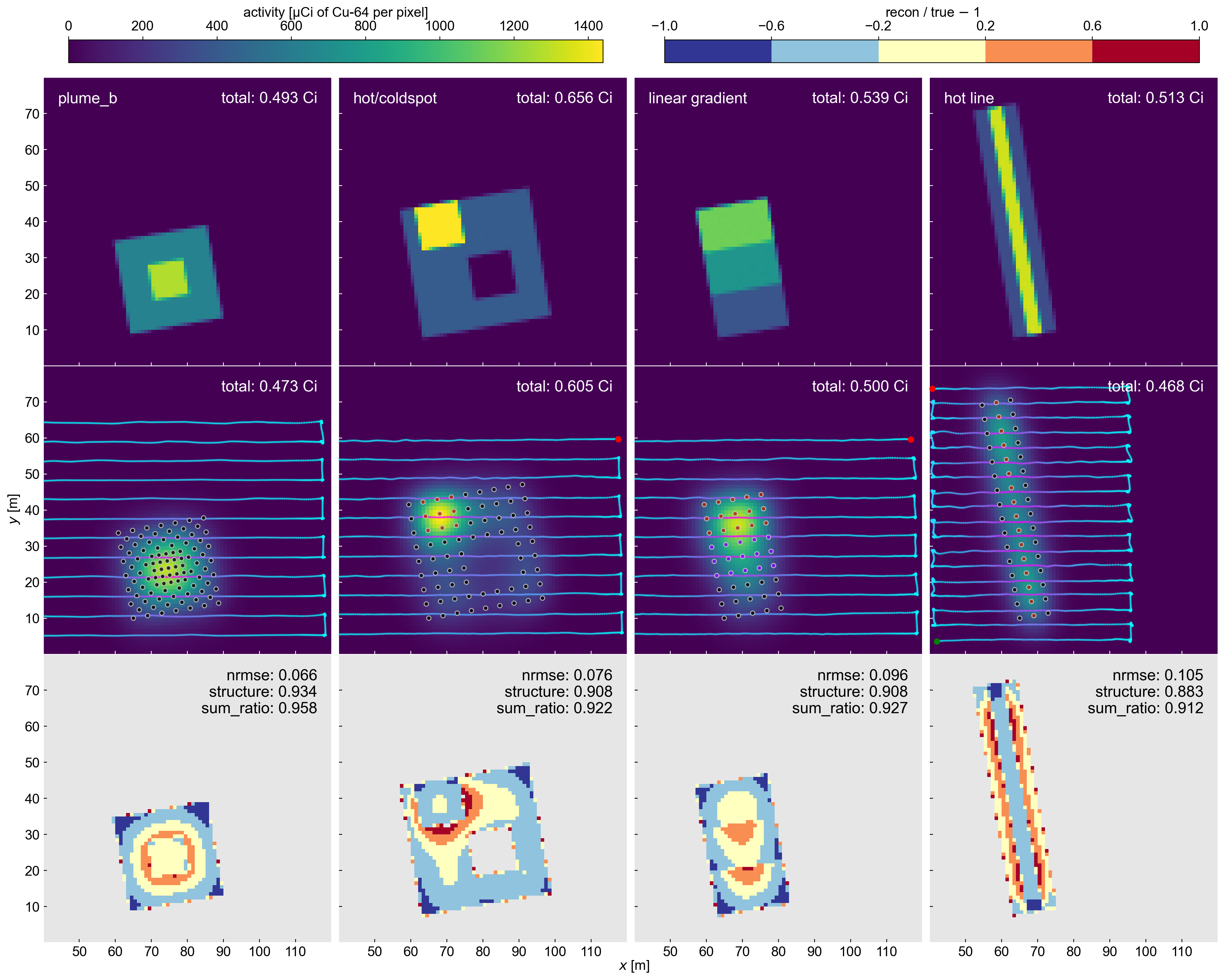}
    \caption{
        Like Fig.~\ref{fig:imaging_study_low}, but for the other four source distributions, each of which featured areas of heightened source activity.
        Left to right: the plume, the hot/coldspot, the linear gradient, and the hot line sources.
    }
    \label{fig:imaging_study_high}
\end{figure*}

Figs.~\ref{fig:imaging_study_low} and~\ref{fig:imaging_study_high} show radiation image reconstruction results from eight different source measurements with the NG-LAMP system at a nominal $6$~m above ground level (AGL), a raster pass spacing of $5.2$~m, and a speed of $2.6$~m/s.
Binmode MAP-EM reconstructions with a time binning of $t_i = 0.2$~s were performed in a 2D image plane of dimensions $125\, \text{m} \times 80\, \text{m}$ with $1\, \text{m} \times 1\, \text{m}$ pixels, and ran in ${\sim}2$~s on a 2019 MacBook Pro using an AMD Radeon Pro 5600M 8~GB GPU\@.
Here we use the sparsity-promoting $\Lhalf{}$ regularizer with coefficient $\beta = 10^{-3}$ and $30$~iterations, which are useful representative values chosen manually based on our prior reconstruction experience.
We note however that we explore the impact of these regularization parameters, as well as many of the other measurement parameters, in the subsequent sections.
We also note that in Figs.~\ref{fig:imaging_study_low}, \ref{fig:imaging_study_high}, and many subsequent image reconstruction figures, the measured trajectories are overlain and colorized by gross count rates (the aforementioned breadcrumbing approach) with a cyan-to-magenta color scale as a qualitative, additional guide, without showing explicit quantitative colorbars.
In general, though, the gross count rates in these figures are ${\sim}1000$~cps far from the source and as high as ${\sim} 15\, 000$~cps above high activity concentration areas and/or sources measured early in the day.
Both trajectories and point source array distributions are shown when their inclusion is relevant and they can be overlain without compromising the clarity of the reconstructed image.

In general, the reconstructions accurately map the shape and, after calibration, the magnitude of the ground truth radiation distributions.
Some overall differences in activity distribution are notable, however.
The interior activities of the shapes are generally over-predicted compared to ground truth, while the edges are under-predicted.
In addition, the corners and edges of the reconstructed images are rounded compared to the sharp features of the true images, falling off smoothly over a few meters rather than instantly over a $1$~m pixel.
As a result, the zero-activity corridor is not very sharp in the $8$~m separation reconstruction, though it is better-resolved in the $12$~m separation reconstruction.
Quantitatively, the ratios of total reconstructed activity $R_\text{tot}$ range from $0.903$ ($8$~m separation) to $0.958$ (plume), indicating high precision and a slight ${\sim}5$--$10\%$ bias in the absolute magnitude of the source reconstructions.
The NRMSE ranges from $0.066$ (plume) to $0.206$ ($12$~m separation), indicating fairly good average shape agreement despite the smoother nature of the reconstructed images.
The sources with more concentrated activity tend to have lower NRMSE values due to their larger number of reconstructed zeros agreeing with the ground truth zeros, so it is also valuable to examine the pixel-wise activity difference maps directly.
Finally, the structure coefficient ranges from $0.877$ ($12$~m separation) to $0.934$ (plume), indicating a high level of perceived shape similarity.

\begin{figure}[!htbp]
    \centering
    \includegraphics[width=1.0\linewidth]{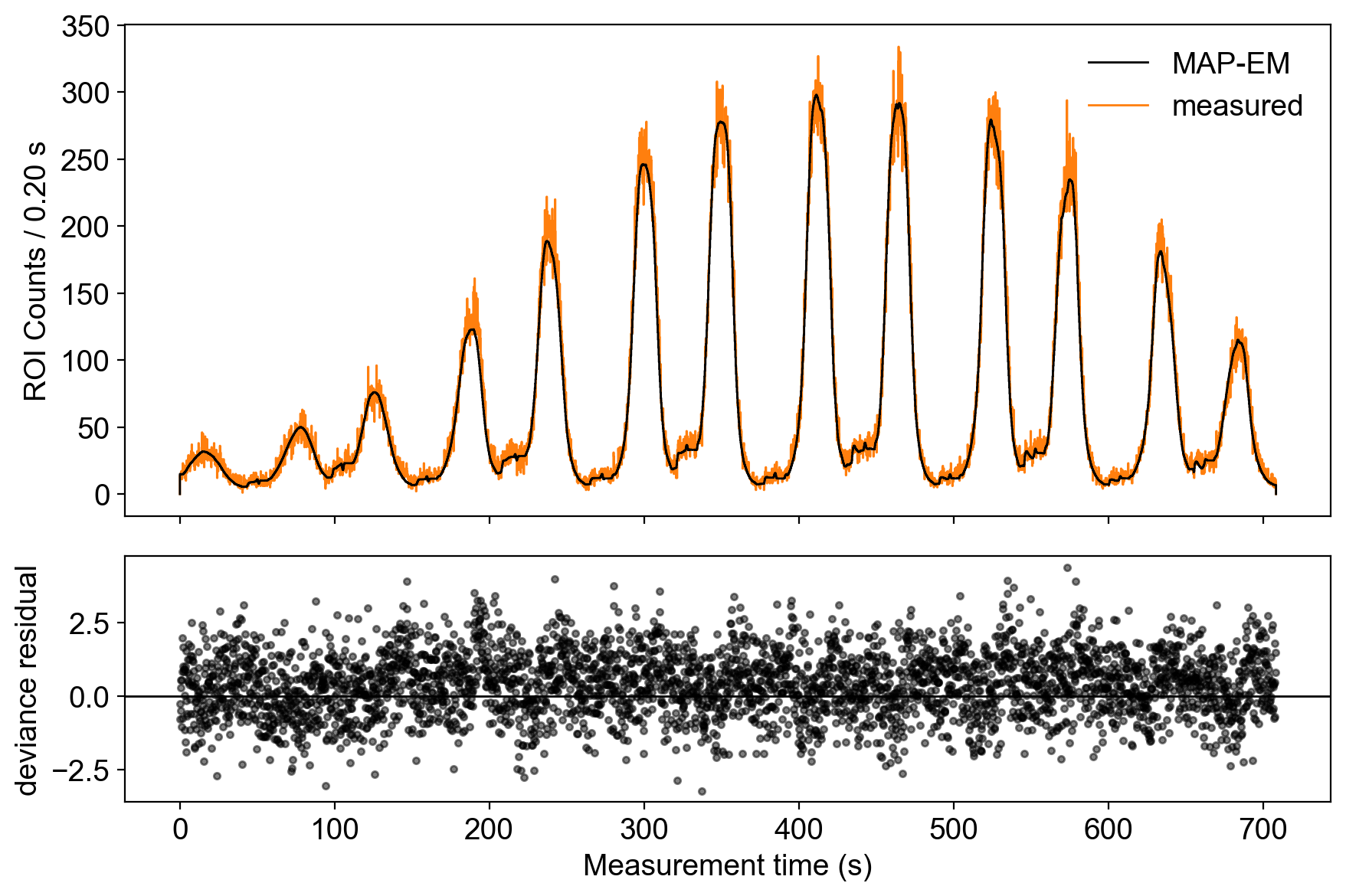}
    \caption{
        Top: MAP-EM-reconstructed (black) and measured (orange) counts in the square source study of Fig.~\ref{fig:imaging_study_low}.
        Bottom: difference between the reconstructed and measured counts in terms of the deviance residual.
    }
    \label{fig:square_study_counts}
\end{figure}

\begin{figure*}[!htbp]
    \centering
    \includegraphics[width=0.9\linewidth]{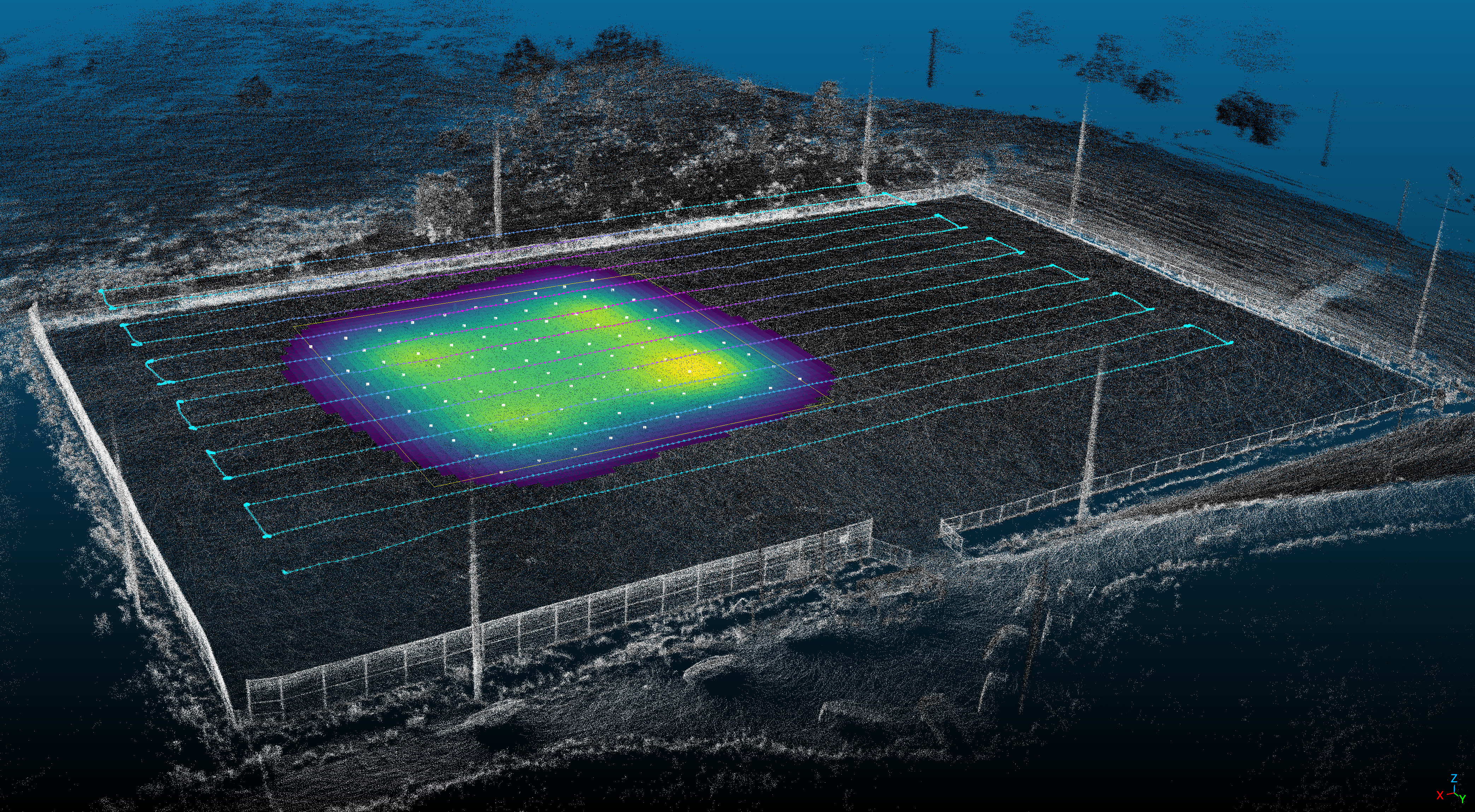}
    \caption{
        3D Scene Data Fusion perspective render of the reconstruction result for the $10 \times 10$ square source study of Figs.~\ref{fig:imaging_study_low} and~\ref{fig:square_study_counts}.
        The point cloud is colorized by the reconstructed image when the image pixel weights exceed an emission rate threshold of $0.1$~s$^{-1}$, and by the lidar intensity otherwise.
    }
    \label{fig:square_study_render}
\end{figure*}

For additional context, Fig.~\ref{fig:square_study_counts} compares the measured and MAP-EM-reconstructed counts vs.\ time from the square source reconstruction of Fig.~\ref{fig:imaging_study_low}.
The deviance residuals and smoothness of the MAP-EM curve qualitatively indicate that the MAP-EM algorithm is not overfitting to noise, and thus that $30$ iterations and $\beta = 10^{-3}$ are suitable choices in this scenario.
For an example energy spectrum, see Fig.~5 of Part~II\@.
Finally, Fig.~\ref{fig:square_study_render} shows a 3D render of the measurement and reconstruction.

\subsection{Altitude study}
\begin{figure}[!htbp]
    \centering
    \includegraphics[width=1.0\columnwidth]{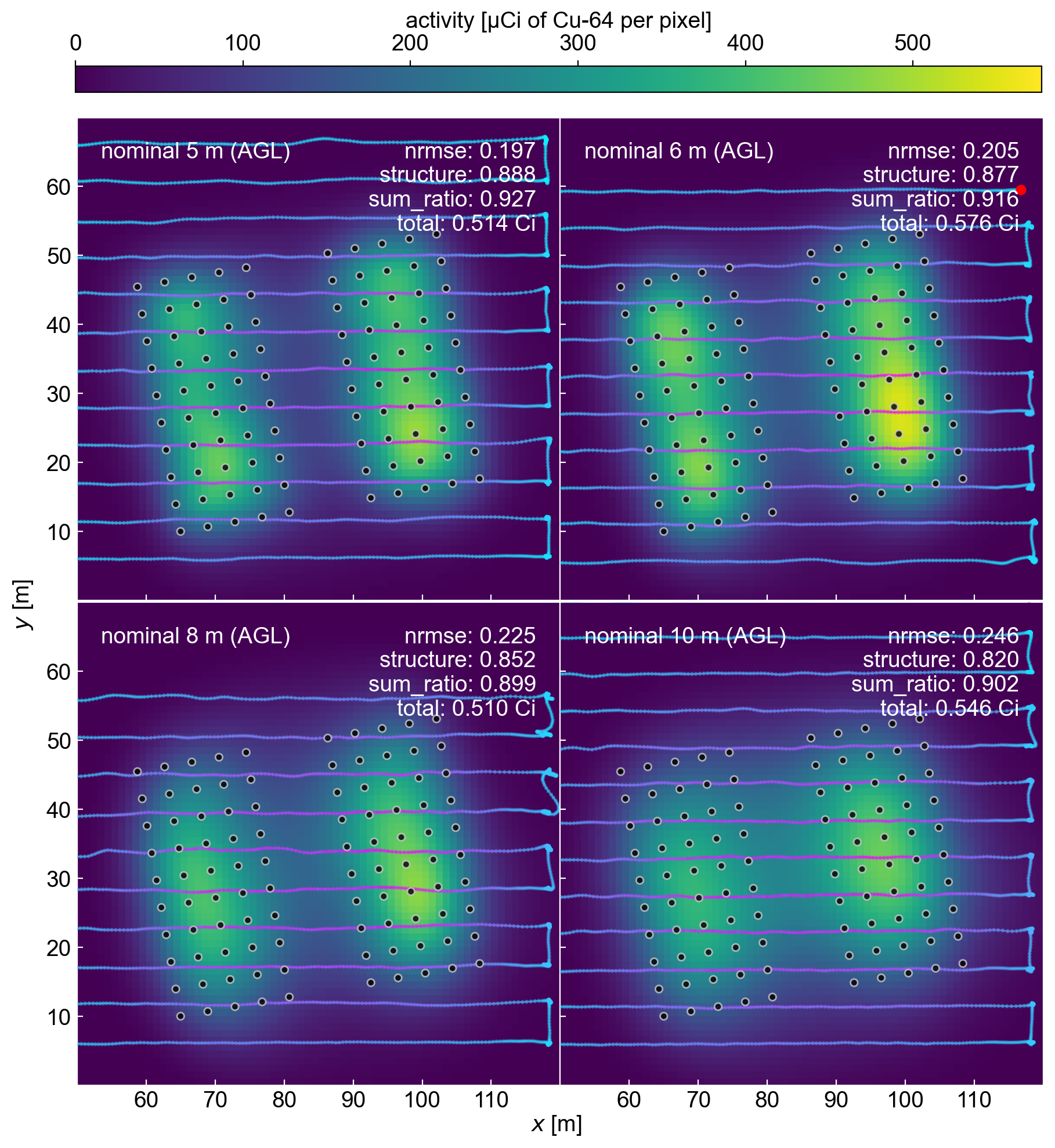}
    \caption{
        Altitude study with the $12$~m separation source.
    }
    \label{fig:altitude_study}
\end{figure}

Fig.~\ref{fig:altitude_study} shows the results from a series of measurements in which the NG-LAMP system was flown over the $12$~m separation source at altitudes of $5$, $6$, $8$, and $10$~m AGL in order to observe trends in the reconstruction vs.~detector altitude.
All three performance metrics degrade monotonically with increasing altitude.
The structure coefficient in particular falls from $0.888$ to $0.820$ and the NRMSE increases from $0.197$ to $0.246$ between $5$~m and $12$~m AGL\@.

These trends suggest that in the idealized case of an omnidirectional radiation detector flying over a flat plane with no intervening material, there is little \textit{imaging} advantage in flying at high altitudes, as opposed to a traditional optical camera where the size of the viewed area increases with altitude.
As shown in the raster spacing study of Section~\ref{sec:raster_spacing_study}, however, non-uniform sensitivity, which can be induced by flying at altitudes much lower than the raster spacing, can cause reconstruction artifacts.
In the present study, the lowest-tested altitude of $5$~m attains the best image quality metrics, striking a balance between the increased sensitivity of flying lower and the increased uniformity of sensitivity of flying higher.
In real measurement scenarios, there are also \textit{operational} advantages to higher altitudes, in particular, the avoidance of obstacles that may attenuate the radiation signal and pose a collision risk for UASs.

\subsection{Replication study}
\begin{figure}[!htbp]
    \centering
    \includegraphics[width=1.0\columnwidth]{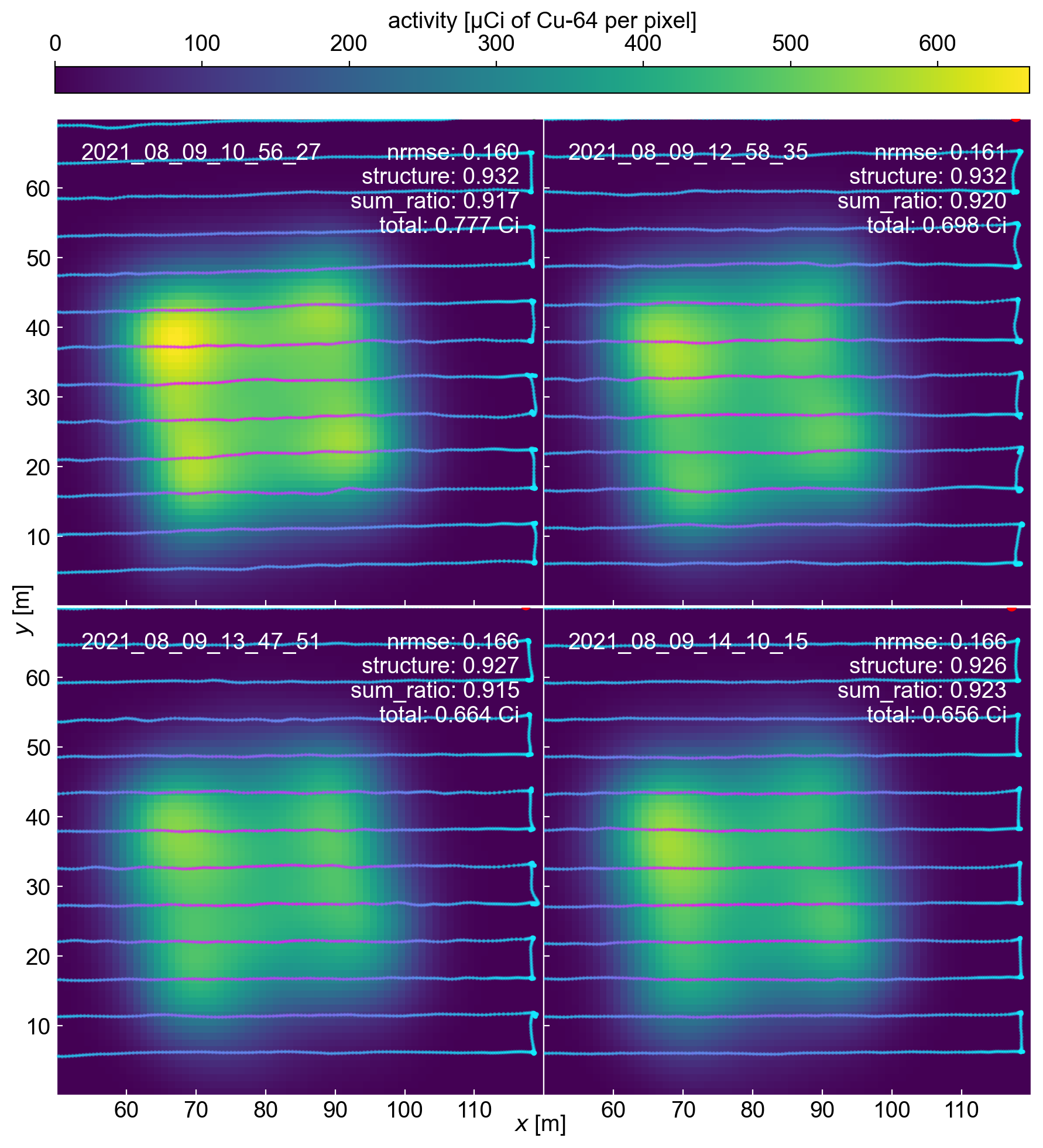}
    \caption{
        Replication study with the square source.
        Note that the top left reconstruction is the same as the square source study of Figs.~\ref{fig:imaging_study_low}, \ref{fig:square_study_counts}, and \ref{fig:square_study_render}.
    }
    \label{fig:replication_study}
\end{figure}

Fig.~\ref{fig:replication_study} demonstrates the replicability of the method, where the NG-LAMP system was flown over the $10 \times 10$~square source at $6$~m AGL in four runs between approximately $1100$ and $1430$~PDT on the same day.
The three performance metrics are remarkably consistent (less than a few percent variation) run-to-run, despite small variations in the measured trajectory and the decay of the Cu-64 sources ($t_{1/2} = 12.7$~hours), both of which are visible in the images.

\subsection{Coarse-graining study}
\begin{figure}[!htbp]
    \centering
    \includegraphics[width=1.0\columnwidth]{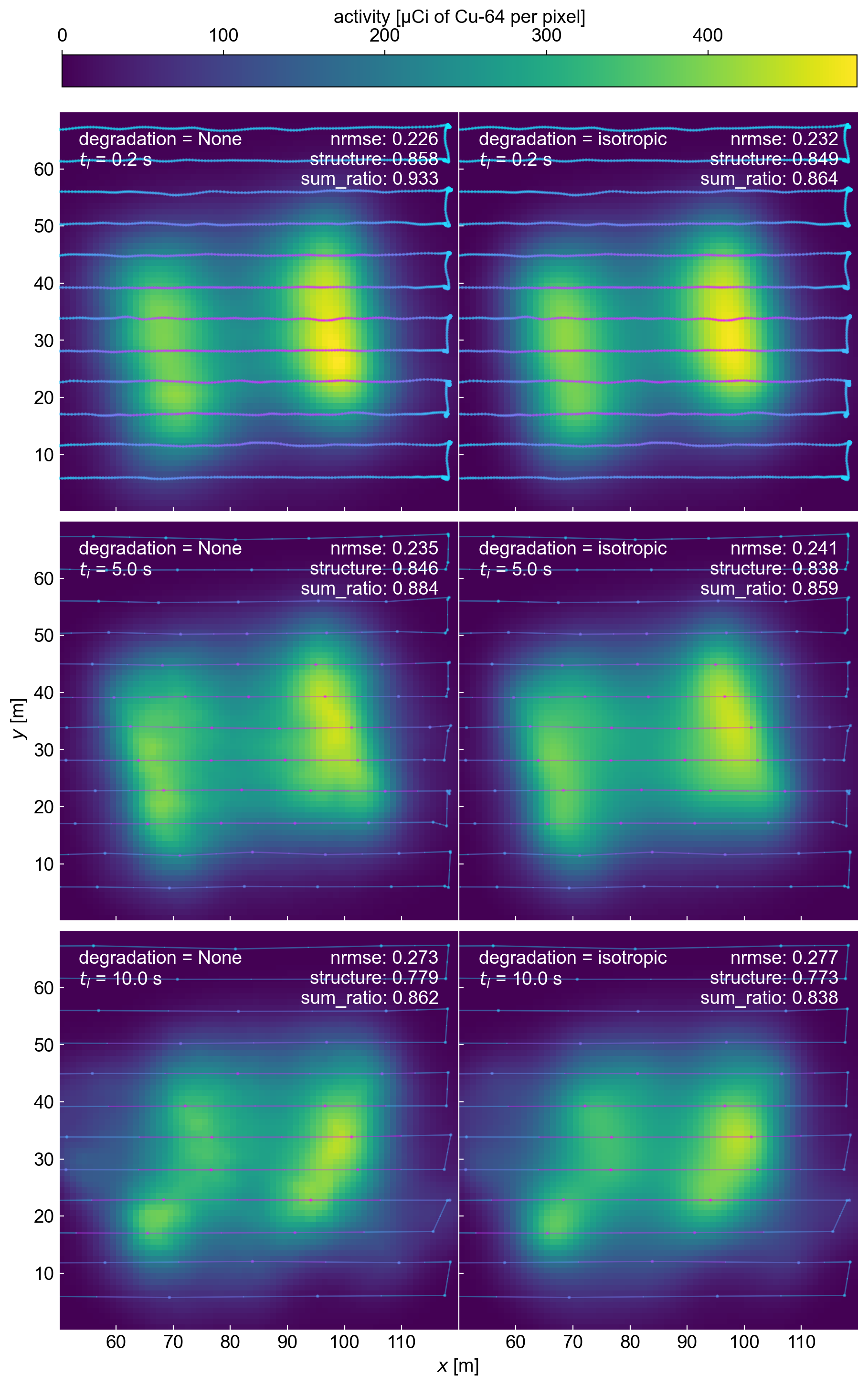}
    \caption{
        Coarse-graining study with the $12$~m-separation source.
        The left column shows image results with increasingly long time-discretization intervals $t_i$ for an analysis of all MiniPRISM detector modules' angular sensitivity~$\eta$ treated individually, while the right column shows the same, but for the modules' efficiencies summed.
    }
    \label{fig:coarse_grain_study}
\end{figure}

Fig.~\ref{fig:coarse_grain_study} shows the results of an analysis of data where the MiniPRISM system was flown over the $12$~m-separation source at a nominal $8$~m AGL and the fidelities of both the data and detector response models were progressively coarsened.
In particular, the data were re-sampled from the original $t_i = 0.2$~s to $t_i = 5$~s and $10$~s dwell times, and the detector response was degraded from its initial $4\pi$ anisotropic multi-crystal response to a single isotropic detector with the same total effective area~$\eta$.
The coarsening of the response to an isotropic detector has only a small negative effect on the reconstruction performance; this is not surprising as the $r^2$ modulation of the counts in Eq.~\ref{eq:sys_mat} is much stronger than the $\eta$ modulation for these airborne surveys.
For additional exploration of the negative effect of response coarsening on reconstruction performance, we refer the reader to Ref.~\cite{vavrek2025demonstration}.
The coarsening of the data to $5$- or $10$-s intervals, conversely, slightly and then significantly degrades reconstruction performance.
While the degradation to $t_i = 5$~s is small according to the three performance metrics, the shape of the distribution noticeably degrades by eye, with the two rectangular activity bands smearing towards the center of the image.
At $t_i = 10$~s, the degradation is even more substantial, with prominent activity smearing (concentrated near the trajectory points) and corresponding changes in the NRMSE and structure metrics.

\subsection{Speed study}
\begin{figure}[!htbp]
    \centering
    \includegraphics[width=1.0\columnwidth]{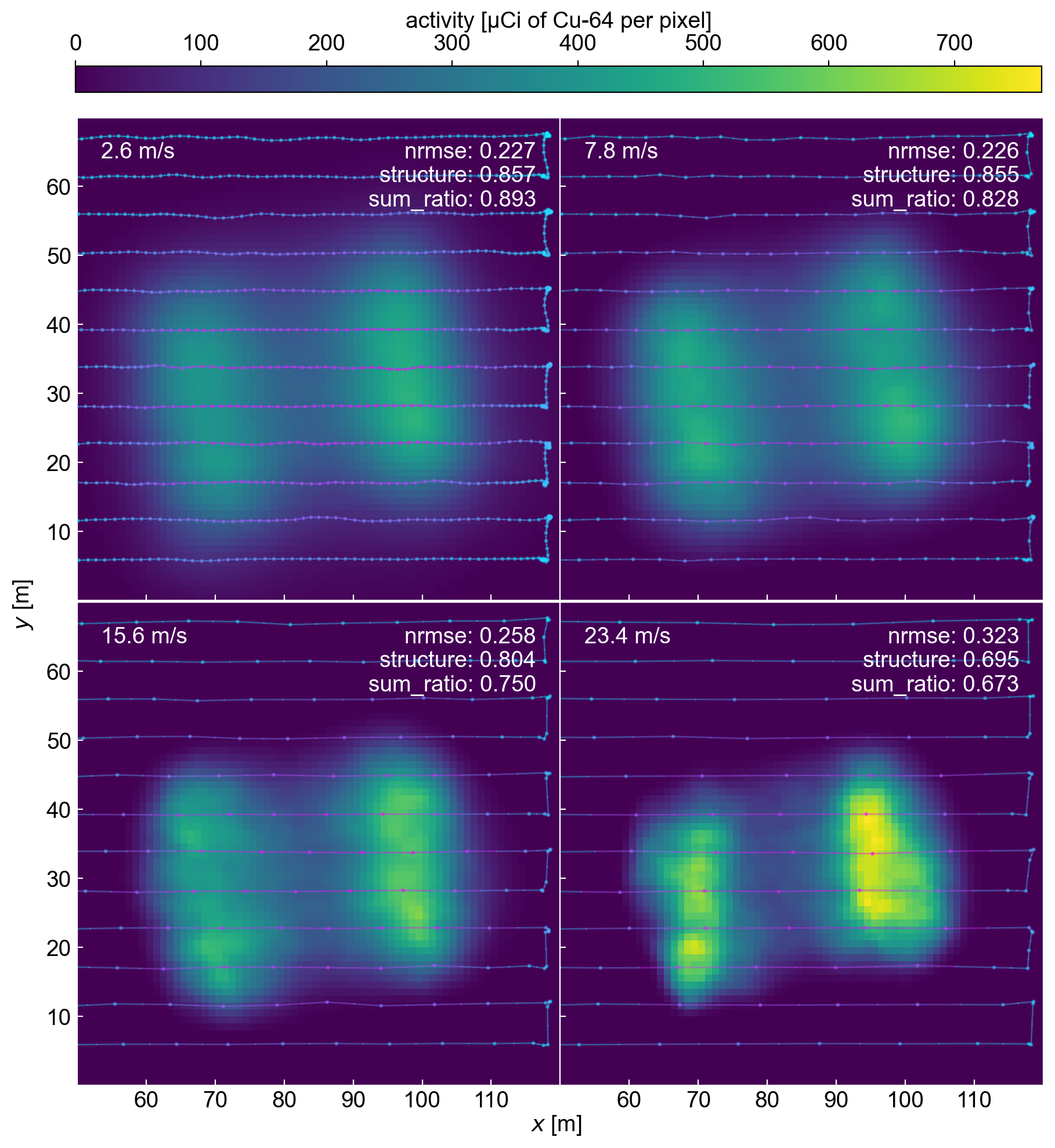}
    \caption{
        Speed study with the $12$~m-separation source.
    }
    \label{fig:speed_study}
\end{figure}

Similar to the coarse-graining study of Fig.~\ref{fig:coarse_grain_study}, Fig.~\ref{fig:speed_study} shows reconstruction results for the $12$~m-separation source with MiniPRISM at $8$~m AGL as the UAS flight speed is increased.
Although during measurements only a single speed of $2.6$~m/s was set, we emulate faster speeds by downsampling the listmode radiation data and compressing the radiation and trajectory timestamps by a constant factor.
In particular, Fig.~\ref{fig:speed_study} shows effective speed increases of $1$, $3$, $6$, and $9\times$, resulting in emulated speeds of $2.6$, $7.8$, $15.6$, and $23.4$~m/s, and corresponding total measurement times of $714$, $238$, $119$, and $79$~s.
Note that we maintain a constant readout interval of $t_i = 0.5$~s, unlike the coarse-graining study, and thus the baseline $1\times$ reconstruction in Fig.~\ref{fig:speed_study} differs slightly from the baseline reconstruction in Fig.~\ref{fig:coarse_grain_study}.

\begin{figure}[!htbp]
    \centering
    \includegraphics[width=1.0\columnwidth]{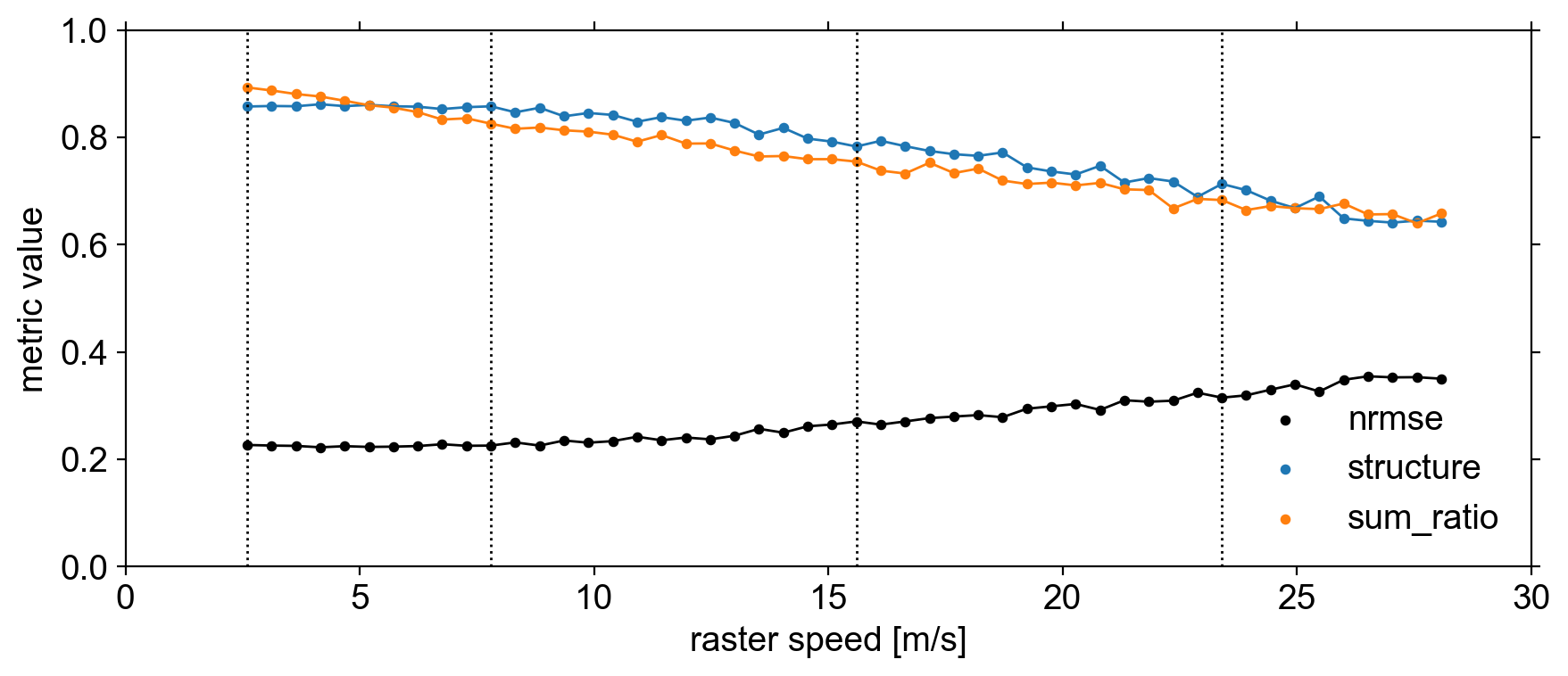}
    \caption{
        Image quality metrics vs.~speed for the $12$~m-separation source.
        The dotted vertical lines correspond to the images in Fig.~\ref{fig:speed_study}.
        Lines between points are drawn only to guide the eye.
    }
    \label{fig:speed_metrics}
\end{figure}

As the speed of the UAS raster increases, the reconstruction noise increases and the image quality decreases due to the reduction in photon statistics and the tendency of the regularizer to concentrate activity around the sparsely-resampled trajectory points, resulting in lower reconstructed activities due to $1/r^2$ effects.
Fig.~\ref{fig:speed_metrics} shows the image quality metrics at more speeds than Fig.~\ref{fig:speed_study}.
While the NRMSE and structure metrics do not markedly degrade at a $3\times$ speedup ($7.8$~m/s), the sum ratio falls from the baseline $0.893$ to $0.828$, and all metrics more noticeably degrade at the $6\times$ speedup ($15.6$~m/s), suggesting a top speed of ${\lesssim} 8$~m/s is suitable for this particular mapping scenario.
Fig.~\ref{fig:speed_metrics} also suggests that the NRMSE and structure metrics are strongly anti-correlated.

\subsection{Regularizer studies}\label{sec:regularizer_studies}
\begin{figure}[!htbp]
    \centering
    \includegraphics[width=1.0\columnwidth]{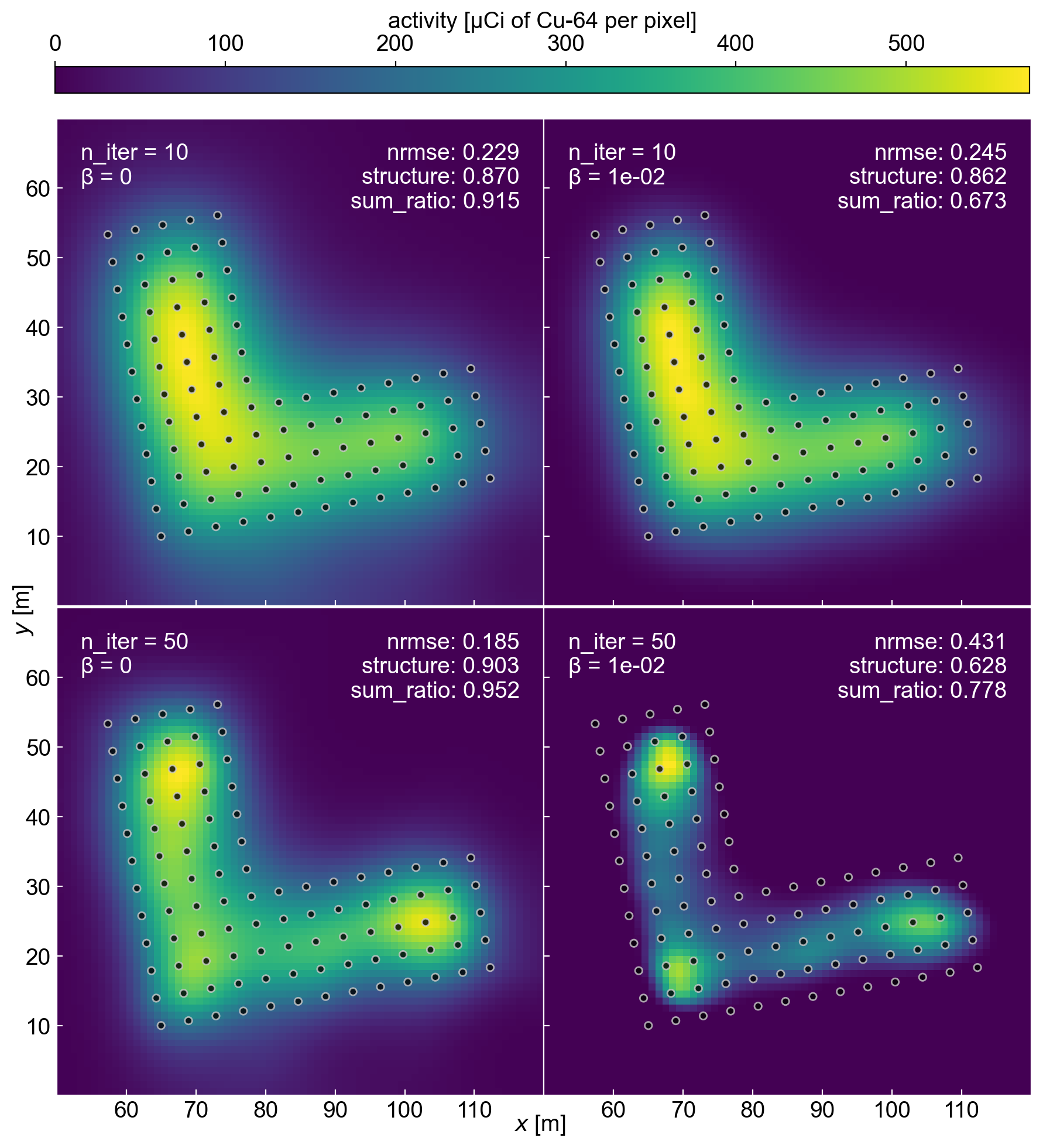}
    \caption{
        \Lhalf{} regularizer study with the \Lshape{} source.
    }
    \label{fig:regularizer0_study}
\end{figure}

Fig.~\ref{fig:regularizer0_study} shows the results of a series of analyses of a single data collection where the NG-LAMP system was flown at $6$~m AGL above the \Lshape{} source, and the two reconstruction hyperparameters were varied between two extreme values.
The \Lhalf{} regularizer coefficient was varied between $0$ (off) and $10^{-2}$ (larger than the $10^{-3}$ used in most other studies) while the number of iterations was varied between $10$ and $50$ (compared to the $30$ used in most other studies).
In all four cases, the overall \Lshape{} is correctly observed, but at various levels of sharpness and thus agreement with the true distribution.
The $(10^{-2},\, 50)$ image is over-regularized, producing a distribution that is significantly smaller than the true source distribution, and has a correspondingly high NRMSE of $0.431$, low structure coefficient of $0.628$, and low total activity ratio of $0.778$.
In particular, the activity is concentrated in internal lobes, resulting in strong underpredictions around the edges. 
At the other end of the spectrum, the $(0,\, 10)$ image has not fully converged, resulting in a modest NRMSE of $0.229$ but an improved structure coefficient of $0.870$, and very smooth edges.
The intermediate case of the $(0,\, 50)$ image performs the best of the four combinations, with an NRMSE of $0.185$, a structure coefficient of $0.903$, and a sum ratio of $0.952$.
This hyperparameter selection performs very similarly to the initial hyperparameters $(10^{-3}, 30)$ used for the \Lshape{} reconstruction in Fig.~\ref{fig:imaging_study_low}, which had an NRMSE of~$0.186$, a structure coefficient of~$0.902$, and a sum ratio of~$0.916$, but the $(0, 50)$ image exhibits faint additional blurring of activity particularly below the \Lshape{} near $x=70$~m.
Therefore we note again that the question of rigorous optimal hyperparameter selection based on comparison to ground truth is of interest, but left for future work.

\begin{figure}[!htbp]
    \centering
    \includegraphics[width=1.0\columnwidth]{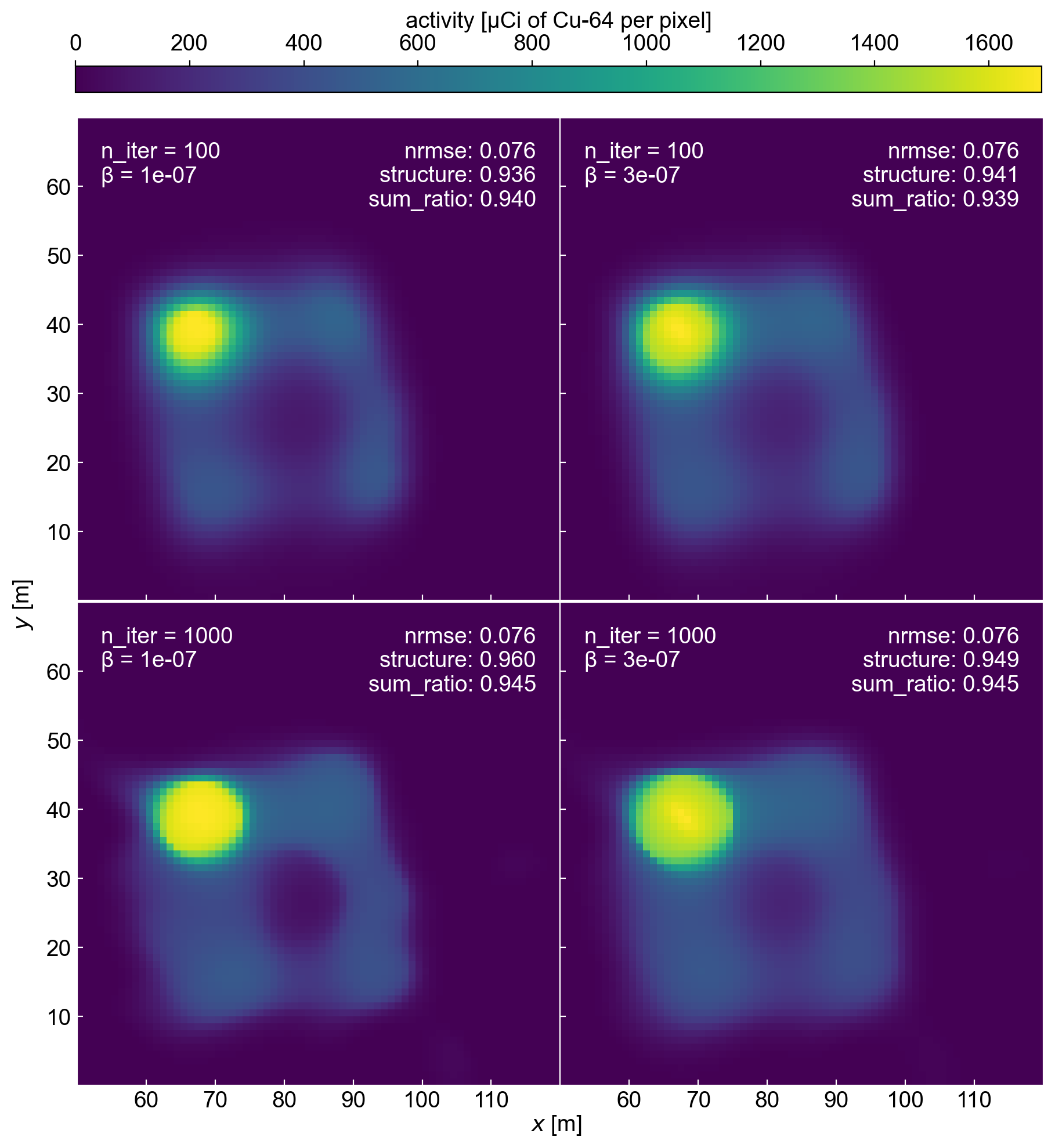}
    \caption{
        TV regularizer study with the hot/coldspot source.
    }
    \label{fig:regularizer1_study}
\end{figure}

Fig.~\ref{fig:regularizer1_study} shows the results of a second regularizer study in which NG-LAMP was flown at $6$~m AGL over the hot/coldspot source, and the TV regularizer coefficient $\beta$ and number of iterations were varied from $10^{-7}$ to $3 \times 10^{-7}$ and $100$ to $1000$, respectively, while using a fixed $\varepsilon = 2.4 \times 10^5$~$\gamma/$s as per Ref.~\cite[\S II-C]{panin1998total}.
The upper bound of the regularizer coefficient was kept lower than the \Lhalf{} regularizer study, and the final two image iterations were averaged together, in order to avoid inducing strong ``checkerboard'' image artifacts that oscillate every iteration at $\beta = 3 \times 10^{-7}$.
Such checkerboard artifacts appear to be a limitation of the TV regularizer at relatively high $\beta$ arising from the fact that minimizing finite differences in $x$ and $y$ smooths the horizontal and vertical bands of the image, but does not constrain diagonally-adjacent pixels~\cite{hu2012higher}.
We also note that for computational tractability, these reconstructions use data summed over all NG-LAMP detector crystals.
Both images using $100$~iterations appear under-converged, with some apparent blurring between the three hot, cold, and baseline regions.
After $1000$~iterations, the hot and coldspots become more distinct from the uniform baseline, with the $\beta = 10^{-7}$ image outperforming the $\beta = 3 \times 10^{-7}$ image in terms of NRMSE and structure coefficient.
In all four TV-regularized images, the summed activity ratio is consistently near $0.94$ and in general closer to $1$ than typical images using the \Lhalf{} regularizer by a few percent.
Similarly the NRMSE values are all close to $0.076$, consistent with the \Lhalf-regularized results for the same dataset in Fig.~\ref{fig:imaging_study_high}, while the structure metrics are better in the TV than the \Lhalf{} reconstructions.
Further discussion of the TV vs.\ \Lhalf{} regularizer performance is given in Section~\ref{sec:discussion}.

\subsection{Raster spacing study}\label{sec:raster_spacing_study}
\begin{figure*}[!htbp]
    \centering
    \includegraphics[width=1.8\columnwidth]{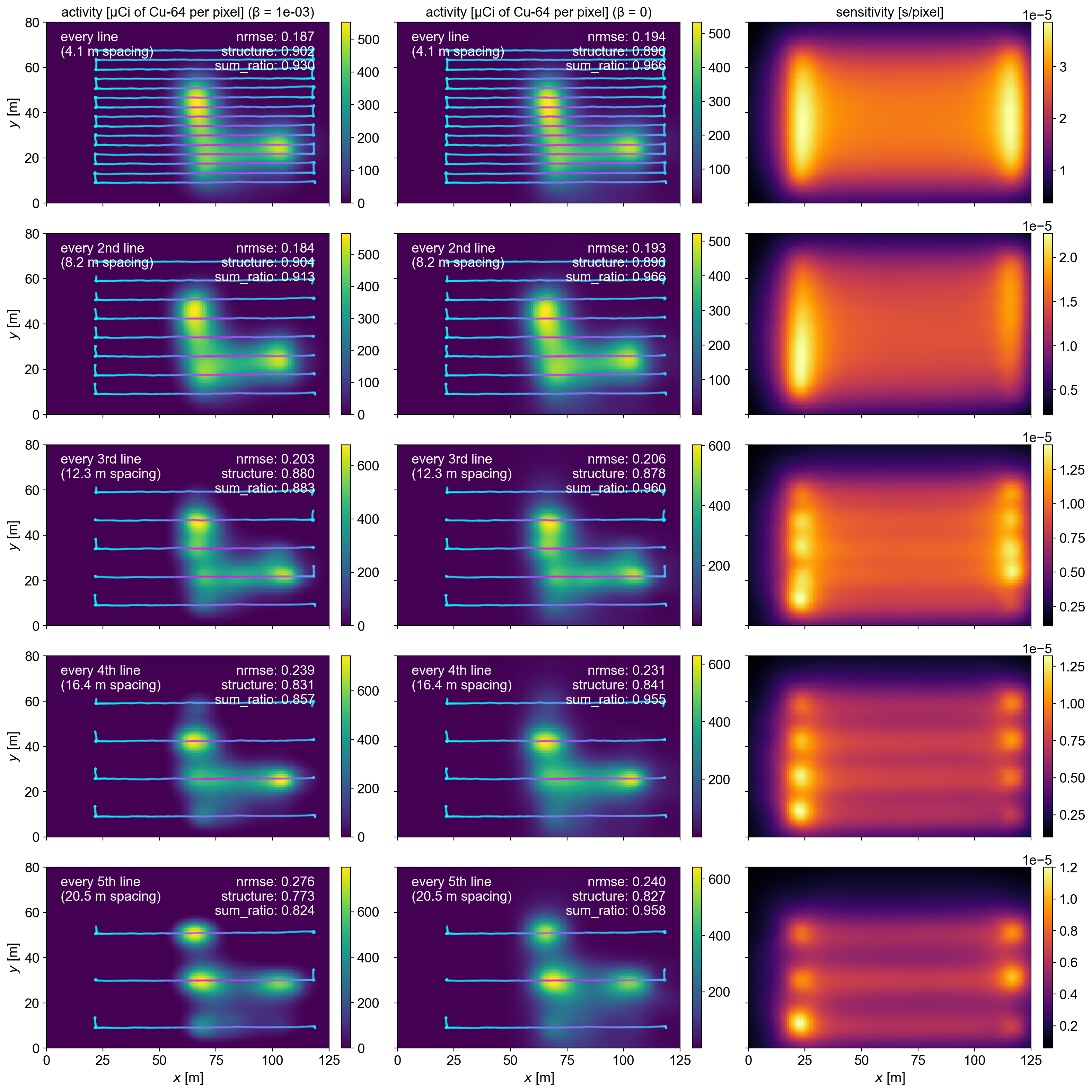}
    \caption{
        Raster spacing study with the \Lshape{} source, showing the reconstructed weights (left and center columns) and sensitivities (right column) when the raster lines are successively cut.
        Note that the color scales vary among rows and columns in order to better highlight the changing shape of the reconstruction and sensitivity maps.
    }
    \label{fig:raster_spacing_study}
\end{figure*}

Fig.~\ref{fig:raster_spacing_study} shows the results of another series of analyses from a single data collection in which NG-LAMP was flown at $6$~m AGL over the \Lshape{} source using a tighter-than-normal line-spacing of $4.1$~m (instead of $5.2$~m) to facilitate testing trajectory cuts in post-processing.
In particular, we cut the trajectory into individual raster lines by simple timestamp cuts and then perform reconstructions (both unregularized and with the standard \Lhalf{} coefficient $\beta = 10^{-3}$) using only every $n\textsuperscript{th}$ raster line for $n=1, \ldots , 5$.

The correct shape of the \Lshape{} source is visible for $n=1,\, 2,\, 3,$ but the shape severely degrades for $n=4,\, 5$.
The degradation is more severe in the \Lhalf{}-regularized reconstructions due to the sparsity-promoting nature of the regularizer.
These trends are borne out by the overall drop in structure coefficient from $0.902$ at $n=1$ to $0.773$ at $n=5$ (regularized) and from $0.896$ to $0.827$ (unregularized), and the corresponding increase in NRMSE from $0.187$ to $0.276$ (regularized) and $0.194$ to $0.240$ (unregularized).
For reference, Fig.~\ref{fig:raster_spacing_study} also shows the sensitivity maps for each post-cut measurement.
The worst-performing images are associated with sensitivity maps that contain noticeable visual banding between the raster lines, indicating that achieving uniform sensitivity over the source area improves quantitative reconstruction.
A similar line of reasoning can also be used to argue that some non-trivial amount of measurement time should be spent flying over source-free areas in order to better capture the contrast between signal and background, though we do not take up this question quantitatively here.

\section{Discussion}\label{sec:discussion}
The results of Section~\ref{sec:results} show good reconstruction quality when measurement parameters and reconstruction hyperparameters are sensibly chosen, typically achieving total activity errors of ${\lesssim}10\%$, quantitative agreement in distribution shape (structure metrics of ${\gtrsim}0.9$ and NRMSEs of ${\sim}5$--$20\%$), and overall agreement with the ground truth by eye.
Now, we discuss how these datasets may be leveraged to further advance quantitative radiation imaging.

Through the altitude, coarse-graining, speed, and raster spacing studies, we have explored whether specific datasets are sufficient for high-quality reconstructions.
While we are able to investigate trends in image quality metrics through these somewhat brute-force parameter sweeps, we expect that information-theory-based approaches may offer additional insights.
An improved understanding of data sufficiency could help both in designing efficient distributed sources experiments, and in determining (perhaps autonomously in real-time) what regions to map and for how long.
Such questions have been explored in the single- or few-point-source case~(e.g., Refs.~\cite{romanchek2021stopping, rolf2020successive}) but to our knowledge have only recently begun to be explored for fully-distributed sources~\cite{shin2022multi}.
Similarly, notions such as the matrix condition number may be useful in exploring the underdeterminedness of the linear model of Eq.~\ref{eq:imaging_model}, though computing the condition number for high-dimensional models remains a computational challenge.

The coarse-graining, speed, and raster spacing studies also demonstrate the capability to resample the dense measured data in post-processing in order to emulate mapping with different detector trajectories.
Because the change in source activity due to radioactive decay during a given measurement is small ($1\%$ over a $10$~minute flight), in future work we may arbitrarily rearrange data in time within the measurement to test, for instance, autonomous trajectory planning methods.

Although we have performed some MAP-EM regularizer studies via Figs.~\ref{fig:regularizer0_study} and \ref{fig:regularizer1_study}, as discussed above we leave a more thorough study of hyperparameter optimization for future work.
As the optimal hyperparameters appear to vary with altitude, system, source, and regularizer type, such an intensive study is beyond the scope of the present paper.
Such a study could also make use of more advanced stopping criteria (e.g., Refs.~\cite{bissantz2008statistical, montgomery2020novel, ben2013heuristic}) compared to the pre-set number of iterations (typically $30$) used here.

Although the TV regularizer may seem like the natural candidate for all source distributions in this work, here we prefer the \Lhalf{} regularizer for several reasons.
First, the \Lhalf{} regularizer uses only one free parameter (the regularization strength $\beta$) while the TV regularizer uses a strength~$\beta$ and a stabilization coefficient~$\varepsilon$.
While Ref.~\cite[\S II-C]{panin1998total} gives a suggestion on how set $\varepsilon$, it depends on the image itself, necessitating a manual iterative tuning process.
Also, as noted in Section~\ref{sec:regularizer_studies}, the TV regularizer can induce checkerboard image artifacts that necessitate a somewhat \textit{ad hoc} averaging of two image iterations.
More generally, it is valuable to explore the \Lhalf{} regularizer since it is expected to improve reconstructions for both plume-like and sharp-edged source distributions, while the TV regularizer will be of limited utility for the former.

Uncertainty quantification (UQ) for these distributed source reconstructions also remains a topic of ongoing work, especially since it will require computationally-tractable high-dimensional UQ methods to compute uncertainties for all $K = 125 \times 80 = 10^4$ image pixels.
The replication study however already indicates a high degree of precision (which suggests a low systematic uncertainty) in reconstructing measurements when holding all parameters constant (other than the decay of the Cu-64 source throughout the day).

Finally, as noted in Part~II~\cite{vavrek2024surrogateII}, in September 2023 we conducted a second distributed sources mapping campaign at the Johns Hopkins University Applied Physics Laboratory (JHU APL) using $100 \times 1$~mCi Cs-137 point-like sources on hilly terrain.
Analysis of this data is underway and performance comparisons between the JHU APL and present WSU campaigns may inform some of the aforementioned questions in the Discussion.
We also anticipate testing Compton reconstruction modes on some of the JHU APL source distributions, though our singles-mode design assumption in Part~I~\cite{vavrek2024surrogateI} will not apply, and thus the sources may no longer appear continuous.

\section{Conclusion}\label{sec:conclusion}

We have performed quantitative reconstructions of surrogate distributed radiological sources as measured by airborne gamma-ray spectrometers.
In general, our reconstructions accurately reconstruct the expected source shapes both by eye and as measured by various quantitative performance metrics.
The absolute activity scale can also be determined to within ${\sim} 10\%$ after applying the activity calibration developed in Part~II\@.
We also performed various parameter sweeps, analyzing reconstruction performance vs.~detector altitude and speed, detector response coarsening, data coarsening, and regularizer type and hyperparameters.
In the $1$-Ci-scale, ${\sim} 1000$-m$^2$-extent examples here, regularized reconstruction performance degraded at altitudes above ${\sim}6$~m, at line spacings above ${\sim}8$~m, or at speeds above ${\sim}8$~m/s, suggesting that keeping UAS flight parameters within these bounds is beneficial for high-quality reconstructions.
Conversely, in this aerial mapping modality, detector response fidelity had a minimal impact on reconstruction performance, with the full anisotropic response models performing only marginally better than their equivalent isotropic models.
We also investigated the impact of two different regularization methods (\Lhalf{} and TV), determined suitable (though not yet optimal) hyperparameters for each, and showed that both are capable of producing high-quality reconstructed images.
More broadly, this work confirms the feasibility of using point source arrays as surrogate distributed sources for distributed source imaging studies, benchmarks LBNL's quantitative SDF capabilities for reconstructing distributed sources, and provides a useful dataset and framework for further studies.

\section*{Acknowledgements}
\small

We thank the students and staff of the Washington State University Nuclear Science Center for their assistance in moving sources and general experimental logistics during the measurement campaign.

This document was prepared as an account of work sponsored by the United States Government. While this document is believed to contain correct information, neither the United States Government nor any agency thereof, nor the Regents of the University of California, nor any of their employees, makes any warranty, express or implied, or assumes any legal responsibility for the accuracy, completeness, or usefulness of any information, apparatus, product, or process disclosed, or represents that its use would not infringe privately owned rights. Reference herein to any specific commercial product, process, or service by its trade name, trademark, manufacturer, or otherwise, does not necessarily constitute or imply its endorsement, recommendation, or favoring by the United States Government or any agency thereof, or the Regents of the University of California. The views and opinions of authors expressed herein do not necessarily state or reflect those of the United States Government or any agency thereof or the Regents of the University of California.

This manuscript has been authored by an author at Lawrence Berkeley National Laboratory under Contract No.~DE-AC02-05CH11231 with the U.S.~Department of Energy. The U.S.~Government retains, and the publisher, by accepting the article for publication, acknowledges, that the U.S.~Government retains a non-exclusive, paid-up, irrevocable, world-wide license to publish or reproduce the published form of this manuscript, or allow others to do so, for U.S.~Government purposes.

This research used the Lawrencium computational cluster resource provided by the IT Division at the Lawrence Berkeley National Laboratory (Supported by the Director, Office of Science, Office of Basic Energy Sciences, of the U.S.\ Department of Energy under Contract No.\ DE-AC02-05CH11231).

\ifCLASSOPTIONcaptionsoff
  \newpage
\fi



%

\bibliographystyle{IEEEtran}
\bibliography{bib}

@misc{joshi2020mfdf,
    title = {{Multi-modal Free-moving Data Fusion (MFDF) v1.0}},
    author = {Joshi, THY and Quiter, BJ and Curtis, J and Bandstra, MS and Cooper, R and Hellfeld, D and Salathe, M and Moran, A and Vavrek, J and Department of Homeland Security and DOD Defense Threat Reduction Agency and USDOE},
    doi = {10.11578/dc.20210701.4},
    url = {https://www.osti.gov/biblio/1806294},
    year = {2020},
}

@techreport{joshi2021radkit,
  title={radkit v1.2},
  author={Joshi, Tenzing and Quiter, Brian and Curtis, Joey and Folsom, Micah and Bandstra, Mark and Abgrall, Nicolas and Cooper, Reynold and Hellfeld, Daniel and Salathe, Marco and Moran, Alex and others},
  year={2021},
  institution={Lawrence Berkeley National Laboratory (LBNL), Berkeley, CA (United States)}
}

@article{Pavlovsky2019,
  title={{3D Gamma-ray and Neutron Mapping in Real-Time with the Localization and Mapping Platform from Unmanned Aerial Systems and Man-Portable Configurations}},
  author={Pavlovsky, R and Cates, JW and Vanderlip, WJ and Joshi, THY and Haefner, A and Suzuki, E and Barnowski, R and Negut, V and Moran, A and Vetter, K and others},
  journal={arXiv:1908.06114},
  year={2019}
}

@inproceedings{pavlovsky2019miniprism,
	address = {Manchester, UK},
	title = {{MiniPRISM}: {3D} {Realtime} {Gamma}-ray {Mapping} from {Small} {Unmanned} {Aerial} {Systems} and {Handheld} {Scenarios}},
	booktitle = {2019 {IEEE} {Nuclear} {Science} {Symposium} and {Medical} {Imaging} {Conference} ({NSS}/{MIC})},
	author = {Pavlovsky, R. T. and Cates, J. W. and Turqueti, M. and Hellfeld, D. and Negut, V. and Moran, A. and Barton, P. J. and Vetter, K. and Quiter, B. J.},
	year = {2019}
}

@article{durrant2006simultaneous,
  title={{Simultaneous localization and mapping: part I}},
  author={Durrant-Whyte, Hugh and Bailey, Tim},
  journal={IEEE robotics \& automation magazine},
  volume={13},
  number={2},
  pages={99--110},
  year={2006},
  publisher={IEEE}
}

@article{bailey2006simultaneous,
  title={{Simultaneous localization and mapping (SLAM): Part II}},
  author={Bailey, Tim and Durrant-Whyte, Hugh},
  journal={IEEE robotics \& automation magazine},
  volume={13},
  number={3},
  pages={108--117},
  year={2006},
  publisher={IEEE}
}

@article{hellfeld2021free,
  title={Free-moving quantitative gamma-ray imaging},
  author={Hellfeld, D and Bandstra, MS and Vavrek, JR and Gunter, DL and Curtis, JC and Salathe, M and Pavlovsky, R and Negut, V and Barton, PJ and Cates, JW and others},
  journal={Scientific reports},
  volume={11},
  number={1},
  pages={1--14},
  year={2021},
  publisher={Nature Publishing Group}
}

@article{shepp1982maximum,
	Author = {L.~A.~Shepp and Y.~Vardi},
	Doi = {10.1109/TMI.1982.4307558},
	Issn = {0278-0062},
	Journal = {IEEE Trans. on Medical Imaging},
	Number = {2},
	Pages = {113-122},
	Title = {{Maximum Likelihood Reconstruction for Emission Tomography}},
	Volume = {1},
	Year = {1982}
}

@article{xu2010lhalf,
  title={{$L_{1/2}$ regularization}},
  author={Xu, ZongBen and Zhang, Hai and Wang, Yao and Chang, XiangYu and Liang, Yong},
  journal={Science China Information Sciences},
  volume={53},
  number={6},
  pages={1159--1169},
  year={2010},
  publisher={Springer}
}

@article{vetter2019advances,
  title={Advances in nuclear radiation sensing: Enabling {3-D} gamma-ray vision},
  author={Vetter, Kai and Barnowski, Ross and Cates, Joshua W and Haefner, Andrew and Joshi, Tenzing HY and Pavlovsky, Ryan and Quiter, Brian J},
  journal={Sensors},
  volume={19},
  number={11},
  pages={2541},
  year={2019},
  publisher={MDPI}
}

@article{zhou2018open3d,
    author    = {Qian-Yi Zhou and Jaesik Park and Vladlen Koltun},
    title     = {{Open3D}: {A} Modern Library for {3D} Data Processing},
    journal   = {arXiv:1801.09847},
    year      = {2018},
}

@article{rudin1992nonlinear,
  title={Nonlinear total variation based noise removal algorithms},
  author={Rudin, Leonid I and Osher, Stanley and Fatemi, Emad},
  journal={Physica D: Nonlinear Phenomena},
  volume={60},
  number={1-4},
  pages={259--268},
  year={1992},
  publisher={Elsevier}
}

@inproceedings{panin1998total,
  title={Total variation regulated {EM} algorithm},
  author={Panin, VY and Zeng, GL and Gullberg, GT},
  booktitle={1998 IEEE Nuclear Science Symposium Conference Record},
  volume={3},
  pages={1562--1566},
  year={1998},
  organization={IEEE}
}

@article{wang2004image,
  title={Image quality assessment: from error visibility to structural similarity},
  author={Wang, Zhou and Bovik, Alan C and Sheikh, Hamid R and Simoncelli, Eero P},
  journal={IEEE transactions on image processing},
  volume={13},
  number={4},
  pages={600--612},
  year={2004},
  publisher={IEEE}
}

@article{vavrek2024surrogateI,
  title={Surrogate Distributed Radiological Sources—Part~{I}: Point-Source Array Design Methods},
  author={Vavrek, Jayson R and Bandstra, Mark S and Hellfeld, Daniel and Quiter, Brian J and Joshi, Tenzing HY},
  journal={IEEE Transactions on Nuclear Science},
  volume={71},
  number={2},
  pages={213--223},
  year={2024},
  publisher={IEEE}
}

@article{vavrek2024surrogateII,
  title={Surrogate Distributed Radiological Sources—Part~{II}: Aerial Measurement Campaign},
  author={Vavrek, Jayson R and Hines, C Corey and Bandstra, Mark S and Hellfeld, Daniel and Heine, Maddison A and Heiden, Zachariah M and Mann, Nick R and Quiter, Brian J and Joshi, Tenzing HY},
  journal={IEEE Transactions on Nuclear Science},
  volume={71},
  number={2},
  pages={224--233},
  year={2024},
  publisher={IEEE}
}

@inproceedings{bissantz2008statistical,
  title={A statistical stopping rule for {MLEM} reconstructions in {PET}},
  author={Bissantz, Nicolai and Mair, Bernard A and Munk, Axel},
  booktitle={2008 IEEE Nuclear Science Symposium Conference Record},
  pages={4198--4200},
  year={2008},
  organization={IEEE}
}

@article{montgomery2020novel,
  title={A novel {MLEM} stopping criterion for unfolding neutron fluence spectra in radiation therapy},
  author={Montgomery, Logan and Landry, Anthony and Al Makdessi, Georges and Mathew, Felix and Kildea, John},
  journal={Nuclear Instruments and Methods in Physics Research Section A: Accelerators, Spectrometers, Detectors and Associated Equipment},
  volume={957},
  pages={163400},
  year={2020},
  publisher={Elsevier}
}

@article{ben2013heuristic,
  title={A heuristic statistical stopping rule for iterative reconstruction in emission tomography},
  author={Ben Bouall{\`e}gue, F and Crouzet, Jean-Francois and Mariano-Goulart, Denis},
  journal={Annals of nuclear medicine},
  volume={27},
  pages={84--95},
  year={2013},
  publisher={Springer}
}

@article{romanchek2021stopping,
  title={Stopping criteria for ending autonomous, single detector radiological source searches},
  author={Romanchek, Gregory R and Abbaszadeh, Shiva},
  journal={{PLOS} One},
  volume={16},
  number={6},
  pages={e0253211},
  year={2021},
  publisher={Public Library of Science San Francisco, CA USA}
}

@article{rolf2020successive,
  title={A successive-elimination approach to adaptive robotic source seeking},
  author={Rolf, Esther and Fridovich-Keil, David and Simchowitz, Max and Recht, Benjamin and Tomlin, Claire},
  journal={IEEE Transactions on Robotics},
  volume={37},
  number={1},
  pages={34--47},
  year={2020},
  publisher={IEEE}
}

@inproceedings{wahl20233d,
  title={{3D} {Compton} Mapping of Point and Distributed Sources with Moving Detectors},
  author={Wahl, CG and Sobota, R and Goodman, D},
  booktitle={2023 IEEE Nuclear Science Symposium, Medical Imaging Conference and International Symposium on Room-Temperature Semiconductor Detectors (NSS MIC RTSD)},
  year={2023},
  organization={IEEE}
}

@inproceedings{macleod20233,
  title={{3-D} Reconstruction of Extended Sources During Dispersal Trial with a {Compton} Imager},
  author={{MacLeod}, A and Murtha, N and Saull, P and Sinclair, L and Andrew, M},
  booktitle={2023 IEEE Nuclear Science Symposium, Medical Imaging Conference and International Symposium on Room-Temperature Semiconductor Detectors (NSS MIC RTSD)},
  year={2023},
  organization={IEEE}
}

@article{daniel2021extended,
  title={Extended sources reconstructions by means of coded mask aperture systems and deep learning algorithm},
  author={Daniel, G and Limousin, O},
  journal={Nuclear Instruments and Methods in Physics Research Section A: Accelerators, Spectrometers, Detectors and Associated Equipment},
  volume={1012},
  pages={165600},
  year={2021},
  publisher={Elsevier}
}

@article{murtha2021tomographic,
  title={Tomographic reconstruction of a spatially-extended source from the perimeter of a restricted-access zone using a {SCoTSS} compton gamma imager},
  author={Murtha, NJ and Sinclair, LE and Saull, PRB and {McCann}, A and {MacLeod}, AML},
  journal={Journal of Environmental Radioactivity},
  volume={240},
  pages={106758},
  year={2021},
  publisher={Elsevier}
}

@article{lange1984reconstruction,
  title={{EM} reconstruction algorithms for emission and transmission tomography},
  author={Lange, Kenneth and Carson, Richard and others},
  journal={J Comput Assist Tomogr},
  volume={8},
  number={2},
  pages={306--16},
  year={1984}
}

@article{shin2022multi,
  title={Multi-Sensor Optimal Motion Planning for Radiological Contamination Surveys by Using Prediction-Difference Maps},
  author={Shin, Tony H and Wakeford, Daniel T and Nowicki, Suzanne F},
  journal={Applied Sciences},
  volume={12},
  number={11},
  pages={5627},
  year={2022},
  publisher={MDPI}
}

@article{vavrek2025demonstration,
  title={Demonstration of a new {CLLBC}-based gamma-and neutron-sensitive free-moving omnidirectional imaging detector},
  author={Vavrek, Jayson R and Pavlovsky, Ryan and Negut, Victor and Hellfeld, Daniel and Joshi, Tenzing HY and Quiter, Brian J and Cates, Joshua W},
  journal={arXiv preprint arXiv:2503.09862},
  year={2025}
}

@article{hu2012higher,
  title={Higher degree total variation ({HDTV}) regularization for image recovery},
  author={Hu, Yue and Jacob, Mathews},
  journal={IEEE Transactions on Image Processing},
  volume={21},
  number={5},
  pages={2559--2571},
  year={2012},
  publisher={IEEE}
}

@misc{girardeau2016cloudcompare,
  title={{CloudCompare}},
  author={Girardeau-Montaut, Daniel and others},
  year={2016},
  note={\url{https://github.com/CloudCompare/CloudCompare}}
}

\end{document}